\documentclass{aa}
\usepackage{graphicx}
\usepackage{txfonts}
\usepackage {csquotes}
\usepackage{gensymb}
\usepackage[utf8]{inputenc}

\title{Precessing ellipses as the building blocks of spiral arms}

\usepackage{amstext}

\begin{document}

\author{M. Harsoula\inst{1}\and
 K. Zouloumi\inst{2}\and 
 C.Efthymiopoulos\inst{3}\and 
 G. Contopoulos\inst{1}}
\institute{$^1$Research center for Astronomy and Applied Mathematics, Academy of Athens,
Soranou Efessiou 4, 115 27 Athens, Greece\\
$^2$Department of Physics,
Section of Astrophysics, Astronomy and Mechanics, University of Athens, Panepistimiopolis, 15784 Athens, Greece\\
$^3$Department of Mathematics, Tullio Levi-Civita, University of Padua, Via Trieste, 63, 35121 Padova, Italy\\
\email{mharsoul@academyofathens.gr;
gcontop@academyofathens.gr; zouloukon@phys.uoa.gr}}  
 
\date{Received ; accepted }

\abstract{
 Stable periodic orbits in spiral galactic models that form families of  precessing ellipses can create  spiral density waves similar to those that are observed in real grand-design galaxies. We study the range in  parameter space for which the amplitude of the spiral perturbation, the  pattern  speed, and the pitch angle collaborate so as to lead to the creation of density waves that are supported by precessing ellipses and their surrounding matter in ordered motion. Quantitative estimates lead to a correlation between the pitch angle and the amplitude of the spiral perturbation and also between the pitch angle and the pattern speed of the spiral arms. These correlations can be regarded as  an orbital analog of a nonlinear dispersion relation in density wave theory.
\vspace{1cm} }

\maketitle

\section{Introduction}

The ubiquity of grand-design bisymmetric spiral arms that are observed in galactic disks in the K band
indicates that the paradigm of the "density wave" remains a valid dynamical model with which the  
spiral structure of many disk galaxies can be described. While the scenario of quasi-stationary spiral density
waves has been strongly criticized  (see Toomre 1977, Athanassoula 1984, Binney
and Tremaine 2008, Sellwood and Calberg 1984, Sellwood 2010 and Dobbs and Baba 2014 for reviews),
the existence of mechanisms regulating the growth rate of spiral density waves, and hence ensuring
their longevity for several pattern periods, remains largely an open question (Bertin 1980, Donner and
Thomasson 1994, see Bertin and Lin 1996 and Contopoulos 2002 for reviews). The density wave
theory fits the description of spiral arms  betten when the spiral amplitude does not
exceed a value of $10\%-20\%$ over a timescale of a few ($\approx$ 5) pattern rotations. In this
regime, nonlinear corrections (Contopoulos 1970, Vandervoort 1971, Norman 1978)
to the linear Lindblad-Lin-Shu theory are required (Lindblad, 1940, 1961, Lin and Shu 1964, 1966).

The basic formulation of density wave theory (see Binney and Tremaine, 2008) deals with spiral
wave perturbations in an axisymmetric disk that is regarded in the framework of either the Boltzmann
equation for collisionless matter or the hydrodynamical equations for gas. On the other
hand, Lindblad (1955) pioneered the orbital description of spiral density waves. In this
description, the central objective is to identify the main families of stellar orbits that
support the spirals, and also to show how these orbits can collaborate to match the
imposed  with the response model of spiral arms (see Contopoulos 1971, Berman and Mark 1977, Vandervoort 1978,
Contopoulos and Grosbol 1986, Patsis et al. 1991). In the orbital description of spiral density
waves, a key element are approximately elliptical stable periodic orbits
whose orientation changes with the distance from the center (parameterized by the Jacobi
energy) so as to produce a response density that closely follows the minima of the imposed
spiral potential. The importance of these periodic orbits as the building blocks of spiral
arms was first stressed by Lindblad (1956, 1957, 1958, 1960, 1961), who called them
``dispersion orbits''. Contopoulos (1970, 1975) developed the theory of resonant periodic orbits
near the inner Linblad resonance, providing the main formulas that allow predicting the
number and stability of the periodic orbits as a function of the Jacobi energy. The resonant
theory of Contopoulos, based on epicyclic action-angle variables, also allows predicting the
structure of the phase space around the periodic orbits, yielding the corresponding
invariant curves as the level curves of a resonant "third integral" (see Contopoulos 2002
for a review). The predictions of this theory were confirmed by Vandervoort  (1973, 1975;
see also Vandervoort and  Monet 1975 and  Monet and Vandervoort 1978), Merzanides (1976) and 
Berry and Smet (1979). These studies provided figures
of the corresponding phase portraits around the elliptical closed orbits computed analytically
(by the third integral) or numerically in simple models of galactic potentials with a
spiral perturbation. On the other hand, Kalnajs (1973) popularized the orbital version of
the density wave theory by providing a schematic figure of the emergence of spiral arms
due to the change in orientation of the axes of the elliptical orbits (``precessing ellipses''),
a figure that was later reproduced in many reviews and books on the subject. Kalnajs (1973) also
computed an approximative formula for the response spiral potential due to the precessing
ellipses, which holds beyond the Wentzel-Kramers-Brillouin (WKB) limit of tightly wound spirals. The overall limit of
the applicability of the precessing ellipses was considered by Contopoulos (1985). The change
in orientation of the elliptical orbits with the energy is explained by resonant
perturbation theory (Contopoulos 1975, 2002, Monet and Vandervoort 1978, see a tutorial
presentation in Efthymiopoulos 2010). Numerical examples of spirals supported by
precessing ellipses have been computed in self-consistent models of spiral galaxies
(Contopoulos and Grosbøl, 1986, Patsis et al., 1991, Patsis and Grosbøl 1996,
Pichardo et al. 2003, Efthymiopoulos 2010, Tsigaridi and Patsis, 2013, Pérez-Villegas
et al. 2015, Chaves-Velasquez et al. 2019).

An important outcome of  linear density wave theory, which is valid for small amplitudes of the spiral perturbation, is the dispersion relation  (see Binney and Tremaine 2008, p. 493), which relates the frequency (pattern speed 
$\Omega_{\rm{sp}}$ of the spiral arms) with the radial wavenumber $k$. For tightly wound m = 2 logarithmic spirals,
the radial wavenumber k and the pitch  
angle $\alpha$ are related by $\alpha= \cot^{-1}(k~r/2)$. In order to understand the relevance of the dispersion 
relation to our results below, we point out that in a given (and fixed) axisymmetric model, the dispersion 
relation depends on the quantities $\kappa(r)$ (epicyclic frequency), $\Omega(r)$ (angular velocity of the circular 
orbit of radius $r$), and $ \Sigma_0(r)$ (axisymmetric disk, as well as on the velocity dispersion 
$ \sigma(r)$ at radius $r$. All these quantities are completely specified by the model that is used to represent the 
axisymmetric component of the disk, that is, by the potential $V_{\rm{ax}}(r)$ and by the form $Q(r)$ of the profile 
of Toomre's $Q$-parameter as a function of the distance from the center (see eq. (15) below). Thus, after an axisymmetric model is fixed, the dispersion relation should be regarded as a relation allowing us to specify the form 
of the function $k(r; \Omega_{\rm{sp}})$, where $k=d\phi_m(r)/dr$ and $\phi_m$ is the phase (position of maxima) 
of the $m-fold$ spiral mode. Hence, solving the differential equation (for $m=2$ spiral arms) $d\phi_2(r)/dr = k(r; \Omega_{\rm{sp}})$, we obtain the form of the phase $ \phi_2(r; \Omega_{\rm{sp}})$ corresponding to the maxima of 
the spiral arms. This implies that within a fixed axisymmetric background (potential + velocity dispersion), 
the linear dispersion relation establishes a correlation between the \textit{form} of the spiral arms (given 
by the function $ \phi_2$) and the value of the spiral 
pattern speed $ \Omega_{\rm{sp}}$. It is important to emphasize 
that in the linear regime, this relation is independent of the amplitude of the spiral perturbation, which is 
simplified altogether from the equations of the linear theory.

However, in nonlinear wave theory, the amplitude of the spiral perturbation $A(r)$ also has to be taken
 into account (see  Contopoulos 1975,1980, Norman 1978). This leads to a form
of the dispersion relation that correlates all three basic parameters of
the spirals, namely, the amplitude $A(r)$, the pitch angle $a,$  and the pattern speed $\Omega_{\rm{sp}}$.

Regarded
from the orbital point of view, two important nonlinear effects are introduced
when the amplitude of the spirals becomes large:\\
i) the `precessing ellipses' (closed orbits of the $x_1$ family)
become largely distorted with respect to the elliptical shape, and\\
ii) chaos is introduced in the system, leading to the lack of a sufficient
number of regular orbits to support the spiral arms.

While these two phenomena have been stressed in previous studies (Contopoulos 1985,
Contopoulos and Grosbol 1986, Patsis et al. 1991, see also Quillen and Minchev 2005),
a systematic study yielding the permissible space in all three parameters
$(A,a,\text{and }\Omega_{\rm{sp}})$ within which the precessing ellipses can support
spiral arms is still lacking, in the literature. This constitutes the
primary motivation for our present study. In particular, we first work with a simple model in which the axisymmetric components (disk and halo)
have parameters with values that are pertinent to a Milky Way model. A bulge is added
to properly secure good kinematic properties of the orbits at all distances across
the disk. We then superpose a $m=2$ logarithmic spiral term to the axisymmetric
potential with a varying amplitude, pitch angle, and pattern speed, and study
the corresponding phase-space structure in the region from the inner Lindblad  resonance (ILR) to
corotation. Our study focuses on the structure of the characteristic curve and the
stability of the $x_1$ family of periodic orbits, along with the extent of
the domain of regular orbits around the $x_1$ family. We also discuss the
role of other families generated near the ILR (in particular, the $x_2$ family),
as well as the range in the parameter values in which the change in pericenter
over galactocentric distance $r$ is such as to support a spiral response
similar to the one imposed by the model. \\
Second, stemming from the basic steps of resonant perturbation theory at the ILR,
a byproduct of our analysis is  to show the convenience of a particular set of variables in
 illustrating all phenomena relevant to the phase-space structure in the region from the 
 IRL and up to the 4:1 resonance. Thus, we depict
all numerical phase portraits (Poincar\'{e} surfaces of section) using
the epicyclic set of canonical variables $(\xi,p_\xi)$, where $\xi =
r-r_c$, $p_\xi = \dot{\xi}$, and $r_c$ is the radius of the circular orbit
of a fixed Jacobi energy $h$ in the flow under the axisymmetric potential
alone. Furthermore, we depict the characteristic curves of closed orbits
($x_1$, $x_2$, etc) showing the quantity
\begin{equation}
S(r_c) = \left(\xi_0^2 + \frac{p_{\xi,0}^2} {\kappa^2(r_c)} \right)^{1/2}
\end{equation}
against $r_c$, where $(\xi_0,p_{\xi,0})$ is the fixed point at which a
closed orbit intersects a surface of section defined by a fixed azimuth. 
This choice of variables (instead of the customary choice,
i.e., $\xi_0$ as a function of $h$) is motivated by the fact that according
to resonant perturbation theory, the relevant quantity characterizing the
precessing ellipses is the amplitude of the epicyclic oscillation for a
particular closed orbit, quantified by $S(r_c)$, as a function of the
average distance of the orbit from the center of the disk. We find that
spiral arms are in general supported by orbits for which $S(r_c)$
is a decaying function of $r_c$ (see Sect. 4).\\
Third, we identify the appearance of  chaos as the main dynamical phenomenon
defining the boundary of the domain in parameter space $(A,a,\text{and }\Omega_{\rm{sp}})$,
outside which spiral arms supported by precessing ellipses cannot exist.
A good criterion for determining this boundary is obtained by computing the
threshold in $(A,a,\text{and }\Omega_{\rm{sp}})$ beyond which the $x_1$ family itself
becomes unstable at distances already close to the ILR. This threshold practically marks the 
nearly complete disappearance of ordered orbits in phase space. Detailed
numerical evidence of these phenomena is given at the end of Sect. 4, leading to 
conclusions that approximately agree with the predictions of  density wave theory, 
and probably also with observations (see the discussion at the end of Sect. 4). 
In particular, this orbital analysis supports the prediction that larger spiral 
amplitudes $A$ as well as slower rotation (smaller $\Omega_{\rm{sp}}$) are consistent 
with more open spirals (higher values of the pitch angle).

The paper is structured as follows. Section 2 presents our model. Section
3 discusses the construction of  spiral density waves using closed orbits. In section 4, an explanation is given of how periodic orbits are determined and how the phase space is constructed. Moreover, the main results of the paper are presented in this section, where we study the role of the amplitude of the spiral perturbation, the pattern speed of the spirals, and the pitch angle in creating realistic spiral density waves. Finally, we summarize our conclusions in section 5.

\section{Model}

We considered a model of a spiral galaxy that contains a combination of an axisymmetric and a spiral potential,
\begin{equation}
V =V_{\rm{ax}}+V_{\rm{sp}}.
\end{equation}
The axisymmetric potential $V_{\rm{ax}}$ is composed of a disk, a bulge, and a halo,
\begin{equation}
V_{\rm{ax}}=V_{\rm{d}}+V_{\rm{b}} +V_{\rm{h}}.
\end{equation}
For the disk potential $V_{\rm{d}}$ , we used a Miyamoto-Nagai model (Miyamoto
and Nagai, 1975) given by the relation 
\begin{equation}
V_{\rm{d}}=\frac{-G M_{\rm{d}}} { \sqrt{ r^{2}+ (a_{\rm{d}}+\sqrt{z^2+b_{\rm{d}}^2})^{2}} }
,\end{equation}
where $M_{\rm{d}}=8.56 \times 10^{10}$ $M_\odot$ is the total mass of the disk, $a_{\rm{d}}=5.3$ kpc and $b_{\rm{d}}=0.25$ kpc. In order to have a 2D disk model,  we took $z=0$ and $r=\sqrt{x^2+y^2}$.
For the bulge, we used a Plummer potential $V_b$ given by the relation
\begin{equation}
V_{\rm{b}}=\frac{-G M_{\rm{b}}}  { \sqrt{ r^{2}+ b^{2}} }
,\end{equation}
where $M_{\rm{b}}=5 \times 10^{10}$  $M_\odot$ is the total mass of the bulge, $r=\sqrt{x^2+y^2}$ and $b=1.5$ kpc.

The halo
potential was a $\gamma$-model (Dehnen 1993) with parameters as in
Pettitt et al. (2014),
\begin{equation}\label{pothalo}
V_{\rm{h}}=\frac{-GM_{\rm{h(r)}}}{r}-\frac{-GM_{\rm{h,0}}}{\gamma r_{\rm{h}}}
\left[-\frac{\gamma}{1+(r/r_{\rm{h}})^\gamma}+\ln(1+\frac{r}{r_{\rm{h}}})^\gamma\right]_r^{r_{h,max}}
,\end{equation}
where $r_{h,max}=100$ kpc, $\gamma=1.02$, and $M_{\rm{h,0}}=10.7 \times 10^{10} M_\odot$,
and $M_{\rm{h(r)}}$ was given by the function:
\begin{equation}\label{mhr}
M_{\rm{h(r)}}=\frac{M_{\rm{h,0}}(r/r_{\rm{h}})^{\gamma+1}}{1+(r/r_{\rm{h}})^\gamma}~~.
\end{equation}
The spiral potential is given by the value $V_{\rm{sp}}$ for $z=0$ of the 3D logarithmic spiral model $V_{\rm{sp}}(r,\phi,z)$ introduced by Cox and Gomez (2002) (see formula (19) in Efthymiopoulos et al. 2020). We have on the disk plane
\begin{equation}
V_{\rm{sp}}= 4 \pi G  h_{\rm{z}} \rho_{0}~ G(r)~\exp \left(- \left(\frac{r- r_{\rm{0}}} { R_{\rm{s}}} \right) \right) {\frac{C}{K B}}~ 
\cos \left[ 2 \left(\varphi-\frac{\ln(r/ r_{0} )}{\tan(\alpha)} \right) \right]
,\end{equation}
where
\begin{equation}
K=\frac{2}{r  |~\sin(\alpha)~|   } , ~~~B= \frac{1+K h_{z}+0.3 (K  h_{z})^{2} }{1+0.3K  h_{z}}
,\end{equation}
and $C=8/(3 \pi)$, $h_{\rm{z}} =0.18$ kpc, $r_0=8$ kpc, $R_{\rm{s}}=7$ kpc, and $a=-13^{\degree}$ was the pitch angle of the spiral arms.  The function $G(r)$ plays the role of a smooth envelope determining the radius  
beyond which the spiral arms are important. We adopted the form 
$G(r)= b-c\arctan((R_{\rm{s0}}-r))$, with $R_{\rm{s0}}=6$~kpc, $b=0.474$, and $c=0.335$. The spiral density was $\rho_{\rm{0}} = 5 \times10^7$, $15 \times10^7$ , or $30 \times10^7~M_\odot~\rm{kpc}^{-3}$ in the three different models under study, respectively.   These three values of the density were chosen so as to yield spiral $F$-strength values consistent with those reported in the literature for a weak intermediate and strong spiral, respectively (see, e.g., Block et al. 2004).

The $F-$strength (Buta et al. 2009) can be defined as either the ratio of the maximum tangential force of the spiral perturbation over the radial force of the axisymmetric background,
\begin{equation}\label{qtan}
F_{\rm{tan}}(r) =\frac{\left\langle F^{\rm{tan}}_{\rm{sp}}(r)\right\rangle_{_{\rm{max}}}}{F_{\rm{r}}(r)}=
\frac{\left\langle\frac{1}{r}\frac{\partial V_{\rm{sp}}}{\partial \varphi}\right\rangle_{\rm{max}}}{\frac{\partial V_{\rm{ax}}}{\partial r}}
,\end{equation}
or the ratio of the maximum total force of the spiral perturbation over the radial force of the axisymmetric background, given by the relation

\begin{equation}\label{qall}
F_{\rm{all}}(r)=\frac{\left\langle F_{\rm{sp}}(r)\right\rangle}{F_{r}(r)}=
\frac{\left\langle\sqrt{\left(\frac{1}{r}\frac{\partial V_{\rm{sp}}}{\partial \varphi}\right)^2 +
\left(\frac{\partial V_{\rm{sp}}}{\partial r}\right)^2}\right\rangle_{\rm{max}}}{\frac{\partial V_{\rm{ax}}}{\partial r}}
.\end{equation}
\begin{figure}
\centering
\includegraphics[scale=0.42]{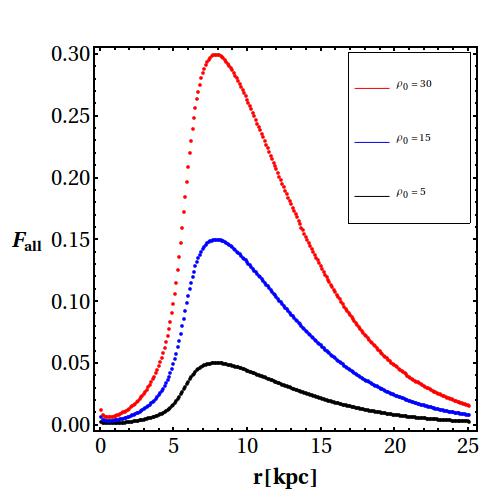}
\caption{Total force perturbation $F_{\rm{all}}$ for different amplitudes of the spiral potential as a function of the radius. } \label{MW}
\end{figure}
Figure 1 shows $F_{\rm{all}}$ as a function of the radius derived from Eq. (11) for three different values of the density $\rho_0$ of Eq. (8), namely $\rho_0=5,15,30 (\times 10^7)~M_\odot~\rm{kpc}^{-3}$. The maximum values of the spiral $F$-strength were $5\%$, $15\%,$ and $30\%,$ respectively. In the intermediate model, the  $F$-strength varies between $5\%$ and $15\%$ in the region $5$ kpc $\leq r \leq 15$ kpc. We note that the observed amplitudes (in  $F-$strength) of the spirals in grand-design galaxies with respect to the axisymmetric background have typical values between $5\%$ and $10\%$ (Patsis et al. 1991, Grosb\o l et al., 2004).\\

\begin{figure}
\centering
\includegraphics[scale=0.32]{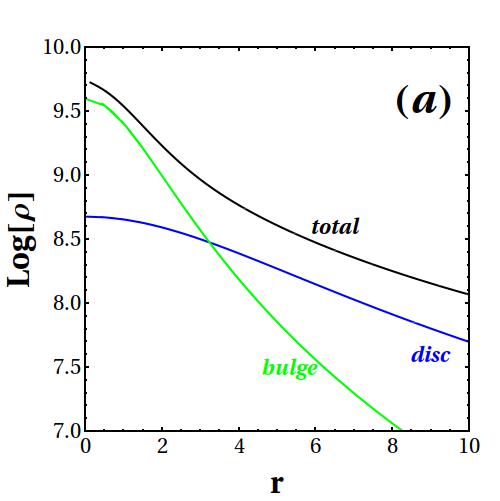}
\includegraphics[scale=0.32]{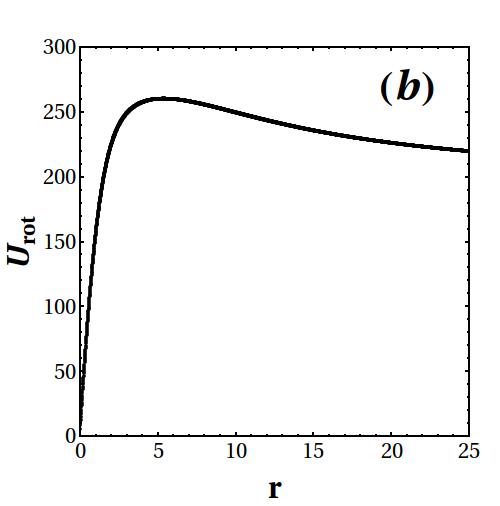}
\caption{(a) Surface density profile of the axisymmetric part of the model. The green curve represents the bulge, the blue curve represents the disk, and the black curve is the total profile. (b) Rotation velocity curve produced by the axisymmetric potential $V_{\rm{ax}}$.} \label{MW}
\end{figure}

Figure 2a shows the surface density profile corresponding to the sum of the disk and bulge component, while Fig.2b shows the
rotation curve produced by the potential $V_{\rm{ax}}$ by the equation: $U_{\rm{rot}}(r)=\sqrt{r ~ \frac{\partial V_{\rm{ax}}(r)}{\partial r}}$.

\begin{figure}
\centering
\includegraphics[scale=0.32]{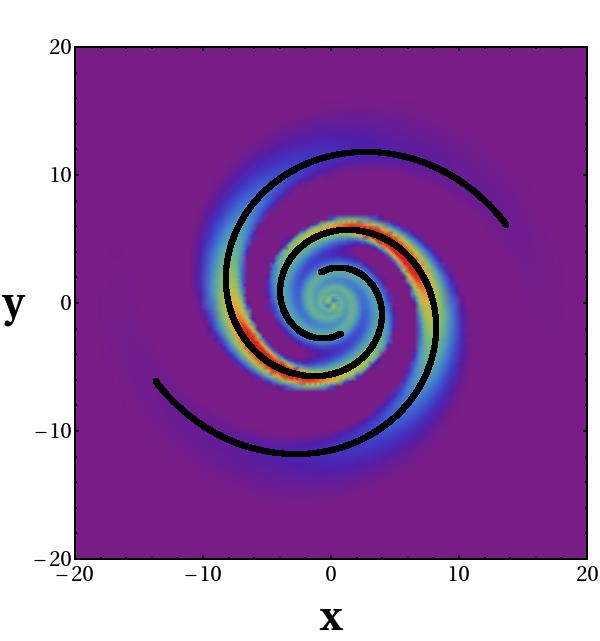}
\caption{Isodensities of the projected surface density of the 3D galactic model (see text). Superimposed are the spiral arms derived from the minima of the spiral potential of Eq. (8).} \label{MW}
\end{figure}
Figure 3 shows an isodensity 
color map of the projected surface density $\sigma(x,y) = \int_{-\infty}^{\infty}
\rho(x,y,z)dz$ in the disk plane, where the density $\rho$ is computed from Poisson's 
equation $\nabla^2V = 4\pi G\rho$ for the 3D potential model $V(r,\phi,z) = V_{\rm{d}}(r) + V_{\rm{b}}(r) + V_{\rm{sp}}(r,\phi,z)$. It is easy to check that the function $\rho(r)$ is positive everywhere with this choice of potential model (see also Pettit et al., 2014 and references within and Efthymiopoulos et al., 2020). The maxima of the density are plotted in red. Superimposed are the spiral arms (black curves) derived from the minima of the spiral potential of Eq. (8), which almost completely coincide with the maxima of the projected density of the galactic model. We note that the phase differences between the minima of the potential and the maxima of the spiral  density have been suggested to be important close to and outside the corotation region (for a review, see Zhang 2018). However, the spirals considered in our work end at the 4:1 resonance (see below), which is reached well before the corotation region.  In the following plots, we therefore use the minima of the potential as indicating the shapes of the imposed spirals for simplicity. 

 When a fixed pattern speed $\Omega_{\rm{sp}}$ of the spirals is assumed, the Hamiltonian of stellar orbits in the disk plane in the rotating frame of reference, in polar coordinates, can be expressed as
 \begin{equation}
 H = \frac{  p_r^{2}}{2 }+ \frac{  p_\varphi^{2}}{2r^2 }- \Omega_{\rm{sp}} p_\varphi + V_{\rm{ax}}(r)+ V_{\rm{sp}}( r,\varphi)
 ,\end{equation}
where $p_r$ and $p_{\varphi}$ are the values of the radial velocity
and angular momentum per unit mass in the rest frame.

The angular velocity $\Omega(r_c)$ of a star that moves in a circular orbit of radius $r_c$ under the action of the axisymmetric potential alone  is given by the relation
\begin{equation}
\Omega(r_c) = \sqrt{\frac{1}{r_c} \frac{d V_{\rm{ax}}(r_c)} {d r_c}}.
\end{equation}
The epicyclic frequency $\kappa(r_c)$   at $r = r_c$ is given by

\begin{equation}
\kappa(r_c)=\sqrt{\frac{d^2V_{\rm{ax}}(r_c)}{dr^2_c}+\frac{3}{r_c} \frac{dV_{\rm{ax}}(r_c)}{dr_c}}
.\end{equation}

\begin{figure}
\centering
\includegraphics[scale=0.28]{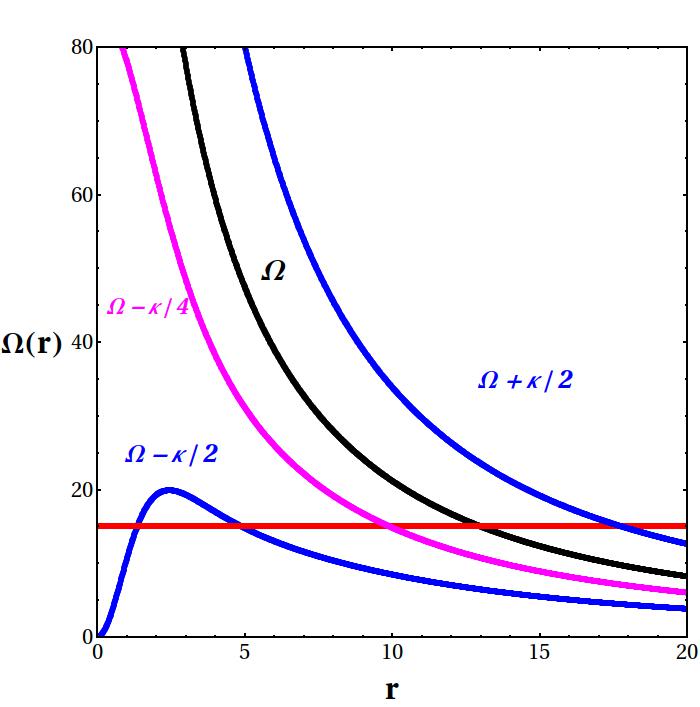}
\caption{Form of the function $\Omega(r)$ and of the resonant combinations of $\Omega(r)$ and $\kappa(r)$, $\Omega-\kappa/2$, $\Omega+\kappa/2$, and $\Omega-\kappa/4$. The selected pattern speed $\Omega_{\rm{sp}}$ determines the radii of the ILR, the 4:1 resonance, and the corotation. } \label{MW}
\end{figure}

Figure 4 shows the function $\Omega(r)$, as well as the combinations $\Omega-\kappa/2$, $\Omega+\kappa/2$ and $\Omega-\kappa/4$.
The section of $\Omega(r)$ with the pattern speed  $\Omega_{\rm{sp}}$ defines the radius of the corotation, while the section of the frequency $\Omega-\kappa/2$ (or $\Omega+\kappa/2$) with the pattern speed  $\Omega_{\rm{sp}}$ defines the radius of the ILR (or the outer Lindblad resonance, OLR). Depending on the model, we may have one or two ILRs. Our model (Fig. 4) has two ILRs, the first and the second ILR, as long as $\Omega_{\rm{sp}}<20 ~\rm{km~ s^{-1}~ kpc^{-1}}$.  Finally, the section of the frequency $\Omega-\kappa/4$
 with the pattern speed  $\Omega_{\rm{sp}}$ defines the radius of the 4:1 resonance. We studied the orbits for three different values of the pattern speed of the spiral potential  $\Omega_{\rm{sp}}$=10, 15, and 20 $\rm{km~s^{-1}~kpc^{-1}}$ (see Sect. 4.2.2).

 \section{Extent of the spiral density waves}

It is well known that spiral density waves cannot extend throughout the entire galactic disk because natural inner and outer barriers limit the extension of these waves.
The spiral density waves are located between the ILR and OLR, but do not reach them (see Dobbs and Baba, 2014 for a review).
The inner natural barrier approximately coincides with the radius of the 2:1 (or ILR) resonance.
The density waves are reflected in this central region before they reach the ILR, and then they are amplified by swing amplification (Goldreich and Tremaine, 1978). Furthermore, the density waves can be  absorbed at the ILR due to Landau damping (Lynden-Bell and Kalnajs, 1972). This absorption of the stellar density waves at the ILR can be avoided, however, if the Toomre $Q$ parameter (Toomre, 1964)
 increases significantly (forming a so-called $Q$-barrier), reflecting the density wave outside the ILR.
 The Toomre $Q$ parameter for a stellar disk is given by the relation
 \begin{equation}
Q=\frac{\kappa \sigma_{\rm{R}}}{3.36 G \Sigma_0}
,\end{equation}
where $\kappa$ is the epicyclic frequency, $\sigma_{\rm{R}}$ is the velocity dispersion, and $\Sigma_0$ is the surface density.
An increase in $Q$ parameter signifies high values of the velocity dispersion. This is  the case inside the ILR also because of the central spheroidal (bulge).  With these assumptions, approximate 'standing-wave' patterns  can exist between a reflecting radius in the inner part of the galaxy and the corotation radius (Bertin et al. 1989).

However, Contopoulos and Grosb\o l, (1986, 1988) have shown that when the amplitude of the spiral arms is strong enough, the outer natural barrier coincides with the appearance of the 4:1 (or ultraharmonic) resonance, which is located inside the corotation. The elliptical periodic orbits become rectangular there and can no longer support the spiral density wave beyond that resonance.
This result was further confirmed in Patsis et al. (1991, 1994, 1997), Lepine et al. (2011), and Junqueira et al. (2013).  Furthermore, Chaves-Velasquez et  al. (2019) showed that the spiral arms of a 3D potential model are
supported by orbits associated with a stable 2D elliptical periodic orbit as well as its vertical bifurcations. The thickness of the spirals supported by such
orbits is compatible with the thickness of the Milky Way disk. However, there are cases where weaker  spirals extend beyond the 4:1 resonance (see, e.g., Grosb\o l and Patsis, 1998).\\
As an overall conclusion, the spiral density waves supported by precessing ellipses should extend in a region starting from slightly outside the ILR and up to the 4:1 resonance. The radii of these resonances are defined by the specific pattern speed. This region is limited in Fig.4 between the functions $\Omega-\kappa/2$ and  $\Omega-\kappa/4$, which is different for different pattern speeds (red line).

 \section{Phase-space structure}
 \subsection{Periodic orbits}

 We now describe the main body of our analysis, which is the study of the phase-space structure and of the orbits supporting the spirals in the region between ILR and corotation in the model of Sect. 2. We chose various values for the spiral amplitude (parameter $\rho_0$ in Eq. (8)), pattern speed $\Omega_{\rm{sp}}$ , and pitch angle $a$. Our study focuses on the form and stability of periodic orbits that support the spiral arms as well as the shape of the phase space around these orbits.

Families of the stable periodic orbits that have shapes of precessing ellipses  can be found by the Hamiltonian  (12), and they correspond to the continuation of the circular orbits of the axisymmetric part of the potential in the region of the  2:1 resonance. The study of such orbits is greatly facilitated using the action-angle variables of epicyclic theory. These are the pair $(\varphi, p_\varphi)$ of Eq. (12), as well as the radial angle and action variables $(\vartheta_r, J_{r})$, defined by

  \begin{equation}
 (r-r_{c})= \sqrt{ \frac{2  J_{r}} {\kappa( r_{c})}  }   \sin ( \vartheta_{r} ),~~~~~~~ p_{r}=\sqrt{2  J_{r} \kappa(r_c)} \cos( \vartheta_{r})
 .\end{equation}
In equation (17), $r_{c}$  represents a radius of a circle in the disk around which we wish to study the form of the phase portrait. The corresponding Jacobi energy $E_j$ is the energy of the circular orbit of radius $r_c$ under the axisymmetric potential $V_{\rm{ax}}$, $H_j(r_c)=H_{\rm{ax}}(r_c)- \Omega_{\rm{sp}} p_{\phi}(r_c)$, where $H_{\rm{ax}}=p_r^2/2+p_{\phi}^2/(2r^2)+V_{\rm{ax}}$ and $p_{\phi}(r_c)=r_c^2 \Omega(r_c)$.

Consider, now, the slow angle $\psi= \vartheta_{r}-2\varphi$, as well as the Poincar\'e canonical variables:
\begin{equation}
\xi= \sqrt{\frac{2 J_{r}}{\kappa(r_c)}} \sin(\psi), ~~~~~~~~~~
P_{\xi}=\sqrt{2 J_{r} \kappa(r_c)} \cos(\psi)
\end{equation}

Using Eqs. (16) and (17), the functions $\xi=\xi(r, p_r,\varphi)$ and $P_{\xi}=P_{\xi}(r, p_r, \varphi)$ are
\begin{equation}
\xi=(r-r_c) \cos(2 \varphi)- \frac{p_r}{\kappa(r_c)} \sin (2 \varphi)
\end{equation}
\begin{equation}
P_{\xi}=p_r \cos(2 \varphi)+ (r-r_c)  \kappa(r_c) \sin (2 \varphi)
.\end{equation}

Equations (18) and (19) can now be used in order to construct a Poincar\'e surface of section $(\xi,P_{\xi})$ for a fixed value of $\varphi$  and a fixed Jacobi constant $E_j$. We chose the value $\varphi=\pi/2,$ and we find a sequence of surfaces of sections in our model defined by the procedure described below.\\
For a specific value of the pattern speed $\Omega_{\rm{sp}}$, we specified a certain value for the radius of the circular orbit $r_c$ and calculated the corresponding angular momentum $p_{\phi_c}=r_c^2 \Omega(r_c)$ and the corresponding Jacobi integral $E_j(r_c)=H_{\rm{ax}}(r_c)-\Omega_{\rm{sp}}$ $p_{\varphi_c}$. For the value of the total Hamiltonian $H=E_j(r_c),$  we then defined various initial
 values  $\xi_0$ and $P_{\xi_0}$ , taking as initial value of $\varphi$ the value $\varphi_0= \pi/2$. Then we calculated the initial values of the coordinates  $r,p_r ,\text{and }p_{\varphi}$ using Eqs. (18) and (19), that is,  $r_0=r_c-\xi_0$ (from Eq. (18)), $p_{r_0}=-P_{\xi_0}$ (from Eq. (19)) and $p_{\varphi_0}=r_0^2 \Omega(r_0)$. Finally, we used  Hamilton's equations
\begin{equation}
\frac{dr}{dt}=\frac{\vartheta H}{\vartheta p_r},~~~~\frac{d\varphi}{dt}=\frac{\vartheta H}{\vartheta p_{\varphi}},~~~~\frac{dp_r}{dt}=-\frac{\vartheta H}{\vartheta r},~~~~\frac{dp_{\varphi}}{dt}=-\frac{\vartheta H}{\vartheta \varphi}
,\end{equation}
with the Hamiltonian (12)
in order to integrate orbits with initial conditions $(\varphi_0,r_0,p_{r_0},\text{and }p_{\varphi_0})$ and find the consecutive iterates ($\xi,P_{\xi}$) on the Poincar\'e section  $\varphi=\pi/2$.

\begin{figure*}
\centering
\includegraphics[scale=0.17]{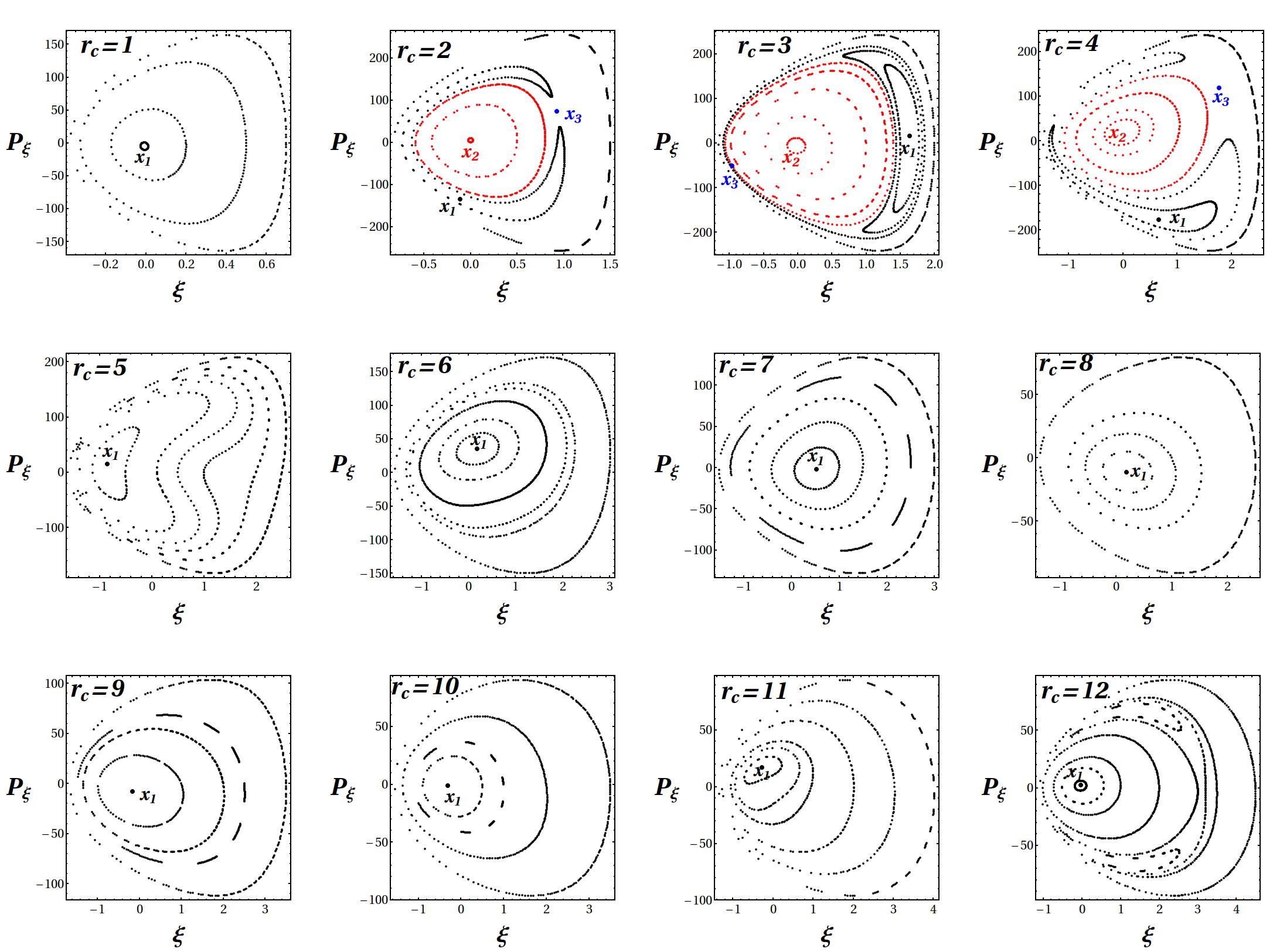}
\caption{Phase-space portraits $(\xi, P_{\xi})$ for the model of Eq. (8) with pattern speed $\Omega_{\rm{sp}}=15~ \rm{km~s^{-1}~kpc^{-1}}$ and density of the spiral potential $\rho_0=5 \times 10^7 M_\odot~\rm{kpc}^{-3}$  for 12  different values of the radius $r_c$ namely $r_c$=1, 2,..., 12 kpc.  Precessing ellipses responsible for the spiral density waves are periodic orbits of the $x_1$ family that correspond to radii between approximately 5 kpc (second ILR) and 11 kpc (4:1 resonance).} \label{MW}
\end{figure*} 
 \begin{figure}
\centering
\includegraphics[scale=0.28]{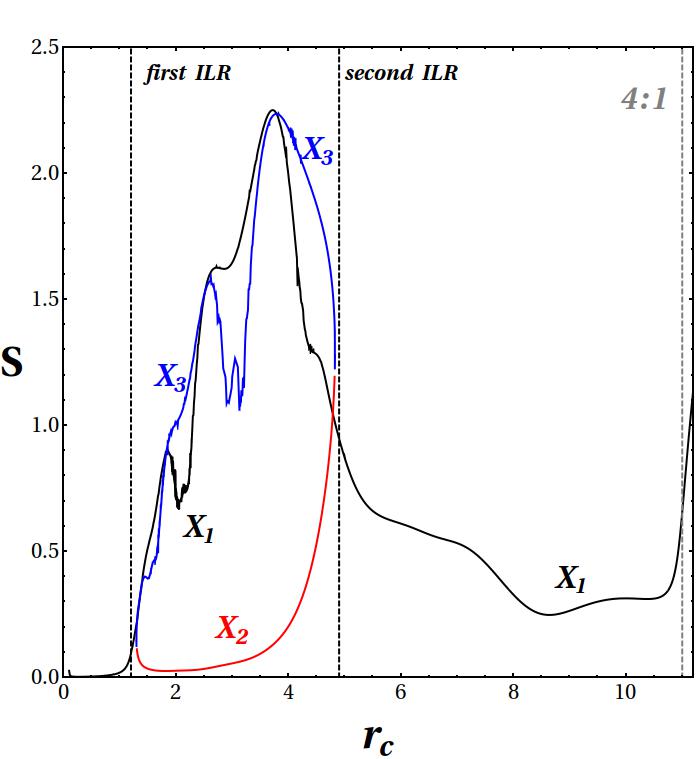}
\caption{Characteristic curves $S(r_c)=\sqrt{\xi^2+p_{\xi}^2/\kappa_c ^2}$ of the periodic families $x_1$ (black), $x_2$ (red), and $x_3$ (blue) for the model of Eq. (8) with $\Omega_{\rm{sp}}=15~\rm{km~s^{-1}~kpc^{-1}}$ and density of the spiral potential $\rho_0=5 \times 10^7 M_\odot~\rm{kpc}^{-3}$.  The dashed black vertical lines correspond to the first and second ILR, and the dashed gray vertical line corresponds to the 4:1 resonance.} \label{MW}
\end{figure}
\begin{figure}
\centering
\includegraphics[scale=0.38]{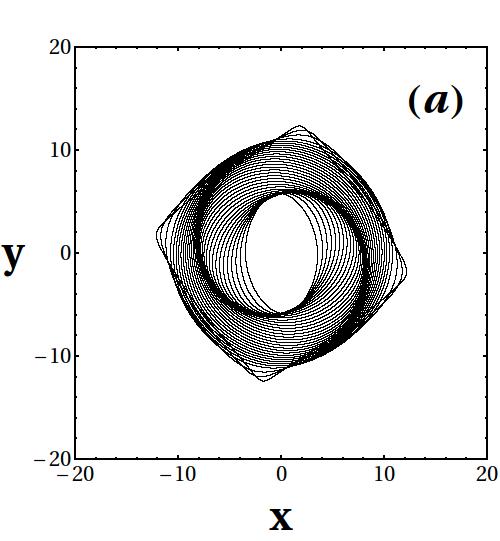}
\includegraphics[scale=0.38]{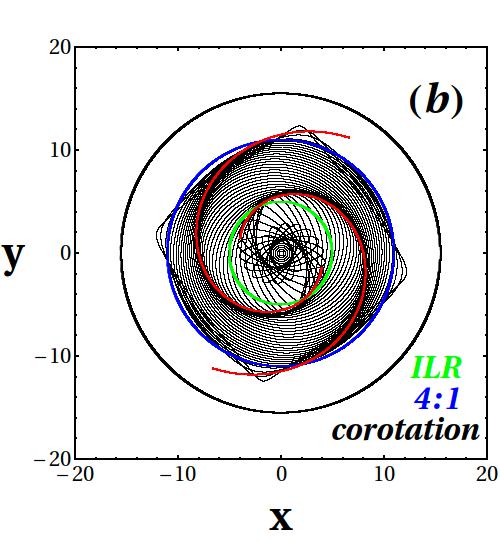}
\caption{ (a) Spiral density waves formed by the precessing ellipses of the elliptical closed orbits of the $x_1$ family, from the model of Eq. (8) for a pattern speed $\Omega_{\rm{sp}}=15 ~\rm{km~s^{-1}~kpc^{-1}}$ and density of the spiral potential  $\rho_0=5 \times 10^7 M_\odot~\rm{kpc}^{-3}$, between the second ILR and the 4:1 resonance. (b) Same as in (a), but with the ellipses of the $x_1$ family inside the second ILR. Superposed are circles corresponding to the second ILR (green circle), 4:1 resonance (blue circle), and  corotation (black circle). The imposed spiral arms (in red) are derived from the minima of the spiral potential of Eq. (8). The coincidence of the imposed spirals with the spiral density wave created by the precessing ellipses is nearly complete. } \label{MW}
\end{figure}
Figure 5 shows the phase portraits (surfaces of section $(\xi, P_{\xi})$) for the parameter $\rho_0=5 \times 10^7 M_\odot~\rm{kpc}^{-3}$ in Eq. (8) and for the pattern speed $\Omega_{\rm{sp}}=15~ \rm{km~ s^{-1}~kpc^{-1}}$.  This value of the pattern speed places the second ILR not far from the inner break of the surface density profile of the axisymmetric part of the potential due to the presence of the bulge (see Fig. 1a). Figure 5 shows the phase portraits  for 12 different values of the radius $r_c$ , namely $r_c$=1, 2, ..., 12 kpc,  spanning a region from the center of the galaxy and up to a radius just outside the 4:1 resonance.

Regarding the main families of periodic orbits in the phase portraits of Fig. 5, we refer to the nomenclature of 
Contopoulos (1975), who called $x_1$, $x_2$ the families of stable and $x_3$ the family of unstable periodic orbits. 
In our model, the precessing ellipses supporting the spirals are the stable periodic orbits of the $x_1$ family.
Four different regions can be distinguished according to the number and stability of periodic orbits: (a)  Inside 
the first ILR, only the $x_1$ family (stable) exists,  (b) between the first and the second ILR, three families exist, namely $x_1$, $x_2$ (stable),  and $x_3$ (unstable). (c) Between the second ILR and the 4:1 resonance, only the $x_1$ family exists (stable or unstable family), and (d)  outside the 4:1 resonance, the $x_1$ 
family still exists, but it no longer supports the spiral arms (see the figures 7,10,13,16).  For $\Omega_{\rm{sp}}=15~ \rm{km~s^{-1}~kpc^{-1}}$, the second ILR and the 4:1 resonance are approximately at $r=5$ kpc and $11$ kpc, respectively (Fig.4). The spirals should thus extend roughly between these two radii.

In the phase portrait of Fig.5, the $x_1$ family is found at the center of the islands of stability, marked 
with black points, while the $x_2$ family is at the center of the islands, marked with red points. The unstable periodic orbit $x_3$ is plotted with a blue dot. The orbits $x_3$ exist only in the region between the first and second ILR, 
and they do not support the imposed spirals (see below). On the other hand, as shown in Fig. 5, the $x_1$ family remains stable at all radii up to the 4:1 resonance. Beyond this resonance, orbits of greater multiplicity bifurcate from the $x_1$ family and substantially affect the structure of the phase space, as in the last panel of Fig.5.

Figure 6 shows the normalized characteristic curves $S(r_c)=\sqrt{\xi^2+P_{\xi}/\kappa_c ^2}$ (where the 
epicyclic frequency $\kappa_c (r_c)$ is given by Eq.(14)) as a function of the radius $r_c$ for the same parameters 
as in Fig. 5. The family $x_1$ is shown in black, $x_2$ in red, and the (unstable) family $x_3$ in blue.  
We note that the $x_2$ and $x_3$ families exist between the first and second ILR. Both families are created 
by a tangent bifurcation close to the first ILR, and then rejoin and disappear by inverse bifurcation close to the 
second ILR. $S(r_c)$  in each case represents the amplitude of the epicyclic oscillation of the corresponding 
(elliptical) periodic orbit. 

A key remark in Fig. 6 is that in the range of the radii where $S(r_c)$ decreases, between the second ILR (at $r_c \approx 5$ kpc) and $ r_c \approx 8.5$ kpc, the ellipses become more circular, and so they do not intersect with each other. The value of $S(r_c)$ reaches a minimum value of about $8.5$ kpc and then increases, first smoothly, and then abruptly. After this latter radius, the orbits become rectangular as they approach the 4:1 resonance. Hence, the $x_1$ orbits start intersecting and no longer support the imposed spirals. As a conclusion, the end of the response spiral density wave should be placed somewhere between the radius $r_c$ corresponding to the minimum of the curve $S(r_c)$ and the  4:1 resonance.

\begin{figure*}
\centering
\includegraphics[scale=0.17]{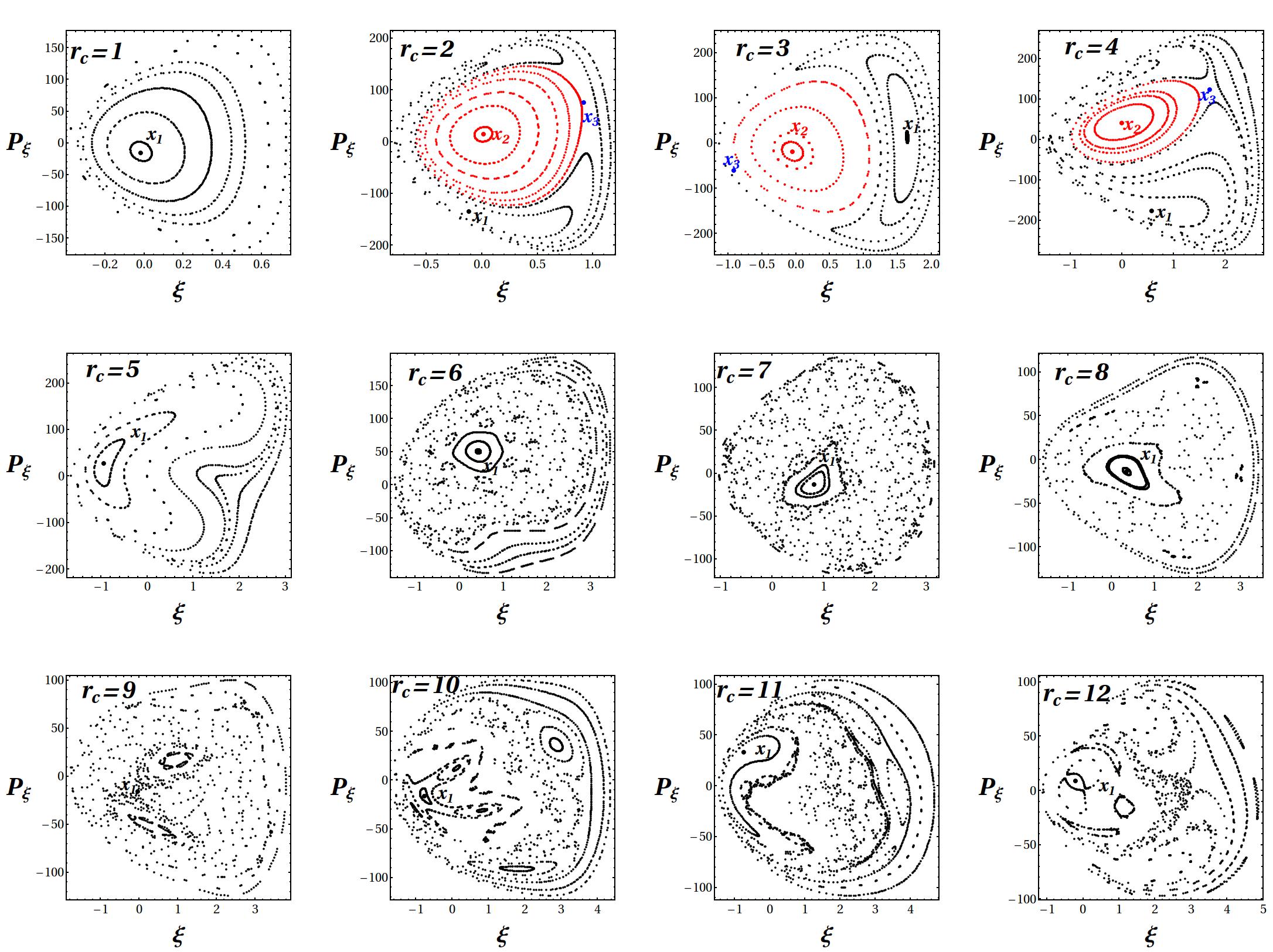}
\caption{Same as Fig.4, but for  $\rho_0=15 \times 10^7 M_\odot~\rm{kpc}^{-3}$.} \label{MW}
\end{figure*}

As shown in Fig. 7, the orbits of the $x_1$ family form precessing ellipses supporting the imposed spirals in the whole region between the second ILR ($\approx 5$ kpc) and a short distance inside the 4:1 resonance. Fig. 7a shows that the $x_1$ family forms a dense and well-defined spiral density wave. In Fig.7b, the $x_1$ family is plotted both inside and outside 
the second ILR. Between the first ILR ($\approx 1.5$ kpc) and the second ILR ($\approx 5$ kpc), the orbits form a more 
fuzzy spiral density wave, while inside the first ILR, the orbits are more circular. In fact, inside the second ILR, the spiral density wave is greatly weakened because the amplitude of the spiral perturbation is close to zero (see Fig. 1). The imposed spiral arms (superposed in the figure in red) are derived from the minima of the spiral potential of Eq. (8). 
The coincidence between imposed spirals and those formed by the elliptical orbits is nearly complete.
 
\subsection{Parametric study} 

The agreement between the imposed spirals (minima of the potential of Eq. (8)) and the response spirals (formed by the elliptical periodic orbits), as well as the regularity of the phase space structure observed in the previous example, holds for a particular choice of parameters ($\rho_0, \Omega_{\rm{sp}}, \text{and }a$). We now study how the above picture is altered by varying independently the amplitude $\rho_0$, the pattern speed $\Omega_{\rm{sp}}$ , or the pitch angle $a$ in the imposed spirals. The results of this parametric study are summarized in Sect. 4.2.3.
 
 \subsubsection{Role of the amplitude of the perturbation}

In order to investigate the role of the spiral amplitude $\rho_0$ (Eq. (8)), we repeated the study of the previous 
subsection for increasing values of $\rho_0$. Figure 8 shows the same phase-space portraits $(\xi, P_{\xi})$ as in Fig.5, but with $\rho_0$ three times larger ($\rho_0=15 \times 10^7 M_\odot~\rm{kpc}^{-3}$). When the amplitude is increased, the main observation is that while the $x_1$ family remains stable for most values of $r_c$, chaos is introduced in the phase space for radii  $r_c>5$ kpc  covering a great part of the phase space around the island of stability corresponding to the $x_1$ periodic orbit. The $x_1$ family itself becomes unstable within a small interval of $r_c$ values (see Fig. 9).
\begin{figure}
\includegraphics[scale=0.28]{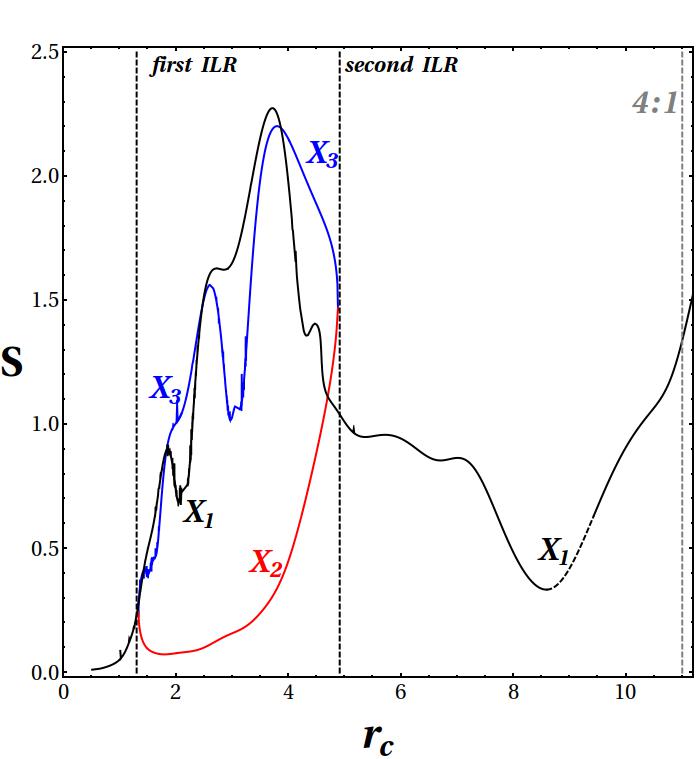}
\centering
\caption{Same as in Fig. 6, but for the  density parameter of the spiral potential $\rho_0=15 \times 10^7 M_\odot~\rm{kpc}^{-3}$. The dashed part of the black curve denotes the unstable periodic orbit $x_1$ .} 
\end{figure}
Figure 9 shows the normalized characteristic curves $S(r_c)=\sqrt{\xi^2+P_{\xi}/\kappa_c ^2}$ in the model with $\rho_0=15 \times 10^7 M_\odot~\rm{kpc}^{-3}$. As in Fig.6, here the $x_2$ and $x_3$ periodic orbits are also created together at a tangent bifurcation close to the first ILR, and then they join and disappear close to the second ILR. 
The value of $S(r_c)$ reaches a minimum value around $8.5$ kpc (same as in Fig.6), and then it increases abruptly. Therefore the end of the response spiral density wave here again should be placed somewhere between the radius $r_c$  corresponding to  the minimum $S(r_c)$ and the radius corresponding to the 4:1 resonance.

Figure 10 shows \textit{the spiral density waves} generated by the $x_1$ family in this model, extending for radii between the second ILR ($r_c \approx 5$ kpc) and  a short distance inside the 4:1 resonance.  The response spirals now appear more concentrated than the spirals in Fig. 7, around the locus of the maximum of the density wave. This is because the forced ellipticity of the $x_1$ orbits increases with $\rho_0$ (see Efthymiopoulos 2010 for a review). In Fig.10b, the $x_1$ family is plotted both inside and outside the second ILR. Here again, the orbits between the first and second ILR (from $\approx 1.5$ kpc to $\approx 5$ kpc) form a more fuzzy spiral density wave, while inside the first ILR, the orbits are more circular.  On the 
other hand, we again observe a nearly complete coincidence between imposed and response spirals beyond the second ILR.
\begin{figure}
\centering
\includegraphics[scale=0.35]{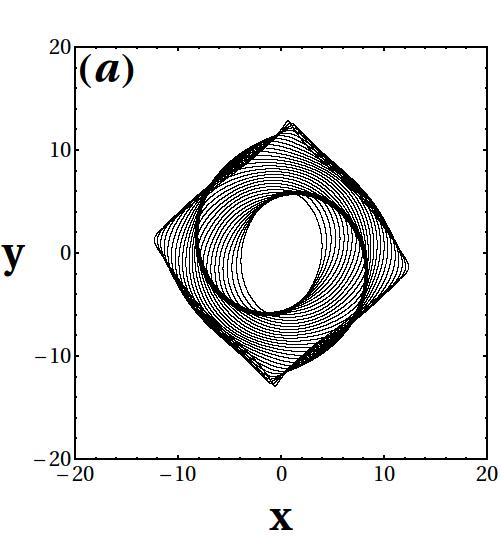}
\includegraphics[scale=0.35]{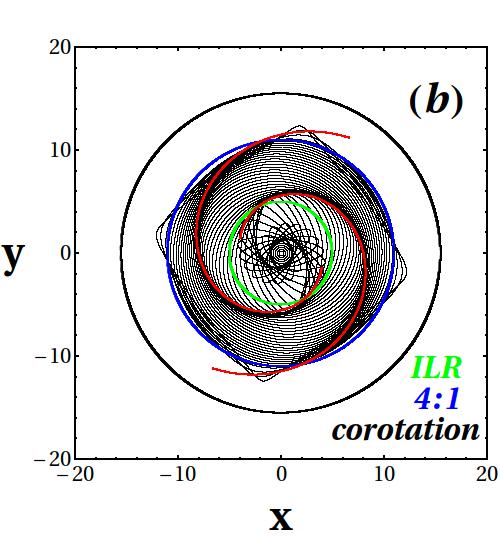}
\caption{ Same as in Fig. 7, but for  $\rho_0=15 \times 10^7 M_\odot~\rm{kpc}^{-3}$. } \label{MW}
\end{figure}
\begin{figure*}
\centering
\includegraphics[scale=0.17]{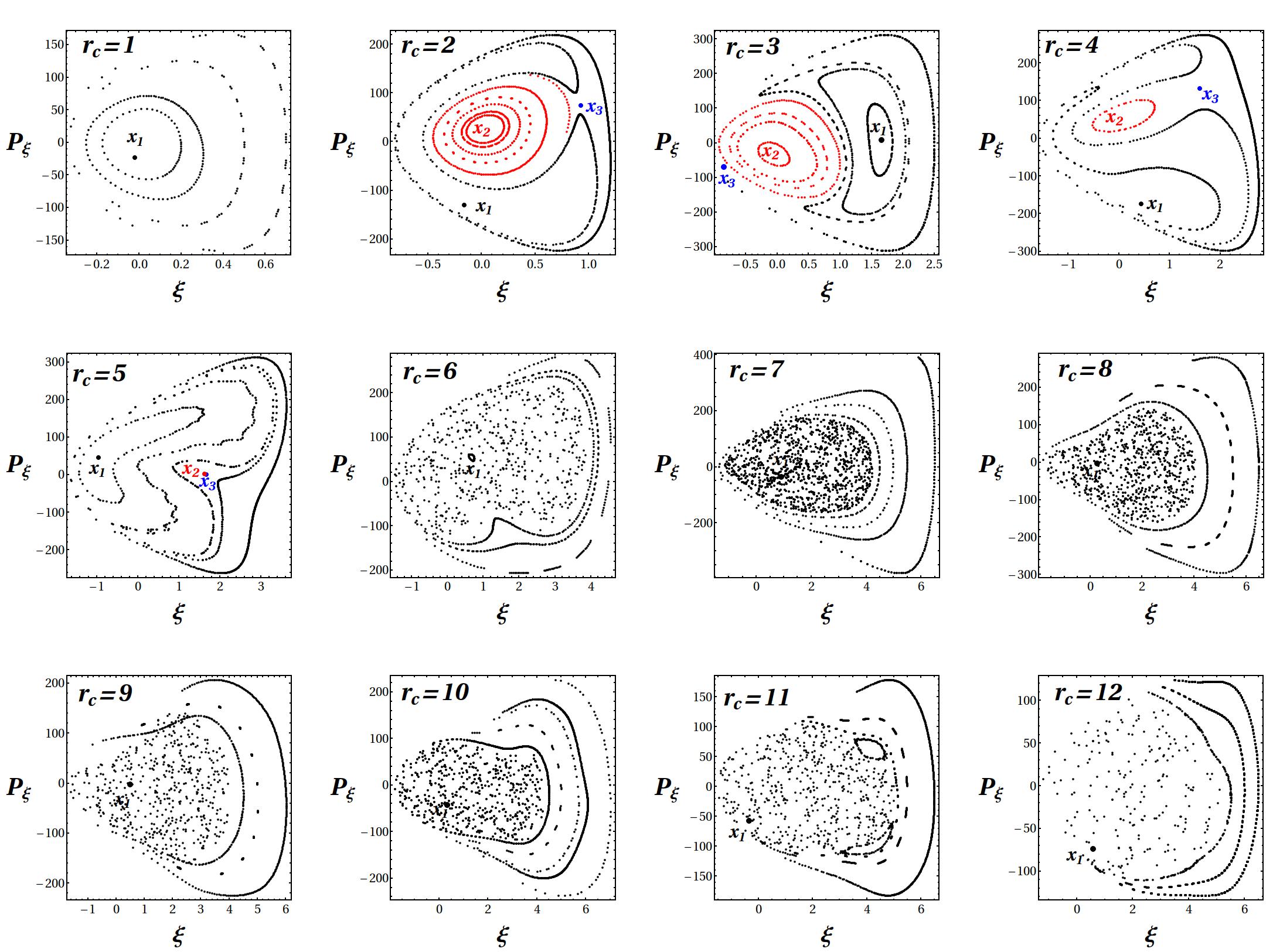}
\caption{Same as Fig.5, but for  $\rho_0=30 \times 10^7 M_\odot~\rm{kpc}^{-3}$.} \label{MW}
\end{figure*}
Figure 11 shows the phase portraits $(\xi, P_{\xi})$ for an even greater (by a factor 6) value of $\rho_0=30 \times 10^7 M_\odot~\rm{kpc}^{-3}$. In comparison with Fig.8, we observe that here chaos is introduced at approximately the same values 
of $r_c$ (i.e., for $r_c > 5$ kpc) as in the previous example. However, the main qualitative difference between these two cases is that in the latter case, the $x_1$ family becomes unstable by a sequence of period-doubling bifurcations starting at $r_c\approx 6.8$ kpc, and there are no ordered orbits around it from there on. As a consequence, no ordered matter is collected around it  that might support a realistic spiral density wave.
 
Figure 12 shows the normalized characteristic curves $S(r_c)= \sqrt{ \xi^2 + P_{\xi} / \kappa_c ^2}$ of this model. The main difference with respect to the previous cases is that the curve $S(r_c)$ forms a nearly constant plateau from the second ILR to the point $r_c \approx 7 \rm{kpc}$ and then marks an abrupt fall to a minimum at $r_c \approx 8$ kpc. As a consequence, the ellipticity of the $x_1$ orbits remains nearly constant in the region between the second ILR and the radius $r_c \approx 7 $kpc. Moreover, when the elliptical orbits of the unstable $x_1$ periodic orbit are plotted for this model (Fig. 13a), the ellipses intersect each other in the whole range of radii and therefore only define fuzzy spiral density waves (compare with Fig. 10). Some secondary spiral arms also appear. Overall, this is an unrealistic spiral density wave that cannot be observed in real galaxies, as there exists no ordered matter around the $x_1$ family. Fig. 13b shows the same information as Figs. 7b and 10b for $\rho_0=30 \times 10^7 M_\odot~\rm{kpc}^{-3}$.

Comparing all three models, we find that the precessing ellipses of the $x_1$ family can support the spirals for amplitudes ($F$-strengths) not exceeding the level $15-20\%$. Beyond this value, the $x_1$ family becomes unstable at a rather small distance  beyond the second ILR ($\Delta r_c \approx 2$ kpc), while chaos dominates the phase space. This is in accordance with estimates of the amplitude of the spiral perturbation of the spiral arms in real grand-design galaxies, which  give a relatively low upper limit ($\approx10-15 \%$) in the forces (Grosb\o l and Patsis 1998, Grosb\o l et al. 2004).
Moreover, long-term evolution of self-gravitating models shows that spiral density waves do not remain viable over many revolutions if the spiral forcing is higher than $5\%$ (Chakrabarti et al., 2003).

For the role of other families of periodic orbits, Fig. 14 shows the precessing ellipses of the $x_2$ family for the model of Eq. (8) for a pattern speed $\Omega_{\rm{sp}}=15~ \rm{km~s^{-1}~kpc^{-1}}$ and a density of the spiral potential $\rho_0=5 \times 10^7 M_\odot~\rm{kpc}^{-3}$ in Fig. 14a,  $\rho_0=15 \times 10^7 M_\odot~\rm{kpc}^{-3}$ in Fig. 14b, and  $\rho_0=30 \times 10^7 M_\odot~\rm{kpc}^{-3}$ in Fig. 14c.  In all three  cases, the response spirals are orthogonal to the response spirals of the $x_1$ family (see Contopoulos 1975). In Figs. 14b and c, they form weak spiral arms in the region between the first and second ILR, where the amplitude of the spiral potential is close to zero (see Fig. 1).

\begin{figure}
\centering
\includegraphics[scale=0.28]{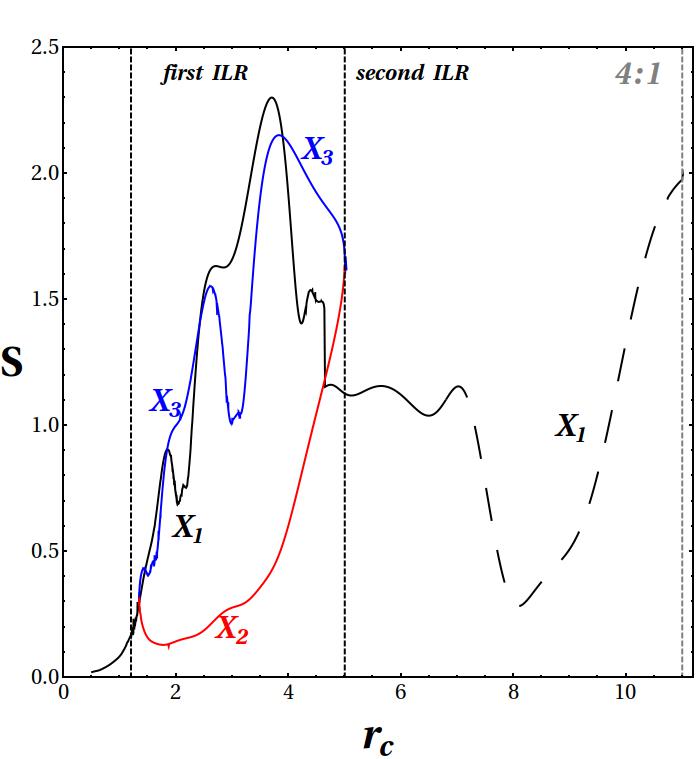}
\caption{Same as in Fig. 6, but for $\rho_0=30 \times 10^7 M_\odot~\rm{kpc}^{-3}$. The dashed part of the black curve denotes that the periodic orbit $x_1$ is unstable.} \label{MW}
\end{figure}

\begin{figure}
\centering
\includegraphics[scale=0.35]{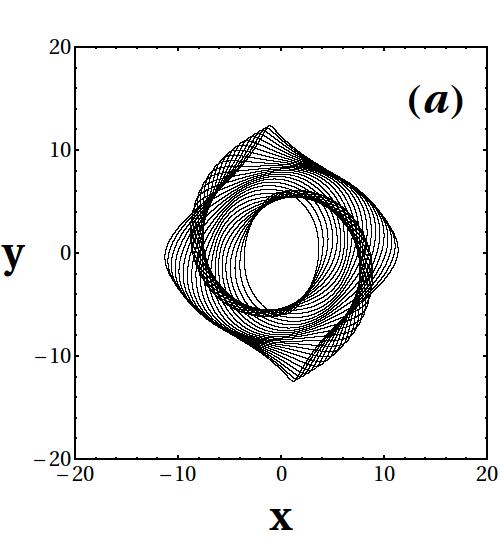}
\includegraphics[scale=0.35]{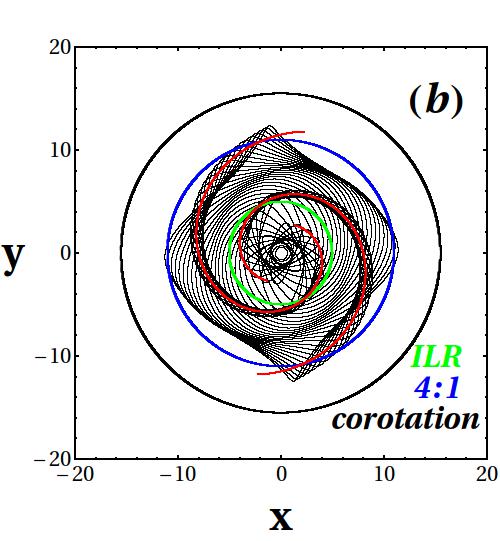}
\caption{Same as in Fig. 7, but for  $\rho_0=30 \times 10^7 M_\odot~\rm{kpc}^{-3}$.} \label{MW}
\end{figure}

\begin{figure*}
\centering
\includegraphics[scale=0.25]{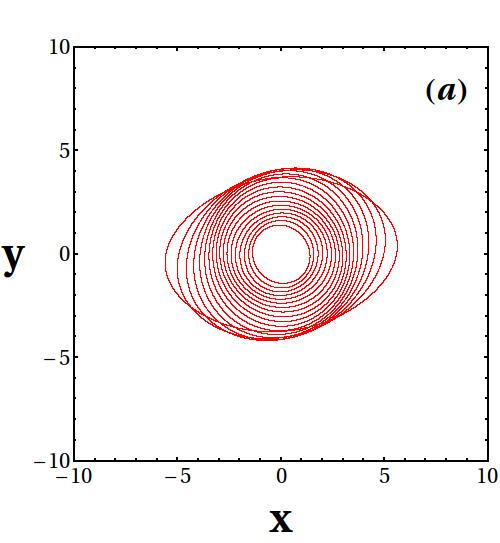}
\includegraphics[scale=0.25]{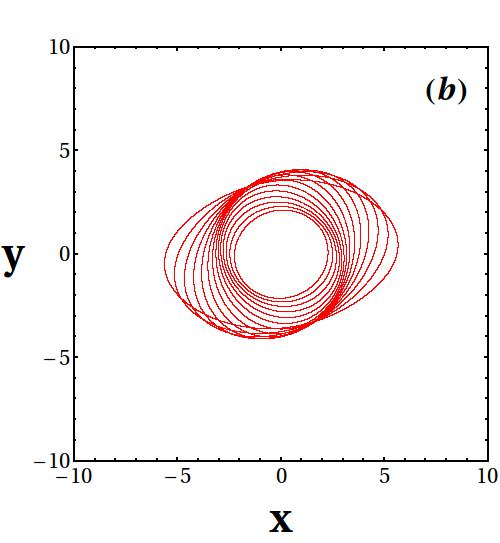}
\includegraphics[scale=0.25]{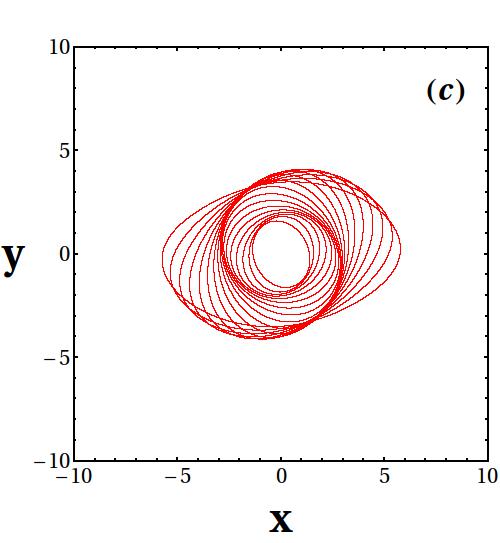}
\caption{ $x_2$ family of orbits for the model of Eq. (8) for (a) a pattern speed $\Omega_{\rm{sp}}$=15$ ~\rm{km~s^{-1}~kpc^{-1}}$ and density of the spiral potential $\rho_0=5 \times 10^7 M_\odot~\rm{kpc}^{-3}$ in (a),  $\rho_0=15 \times 10^7 M_\odot~\rm{kpc}^{-3}$ in (b) and  $\rho_0=30 \times 10^7 M_\odot~\rm{kpc}^{-3}$ in (c).  In all three  cases, these elliptical orbits do not support the spiral wave derived from the galactic potential, and their main axes are perpendicular to the main axes of the $x_1$ family.} \label{MW}
\end{figure*}
\begin{figure}
\centering
\includegraphics[scale=0.40]{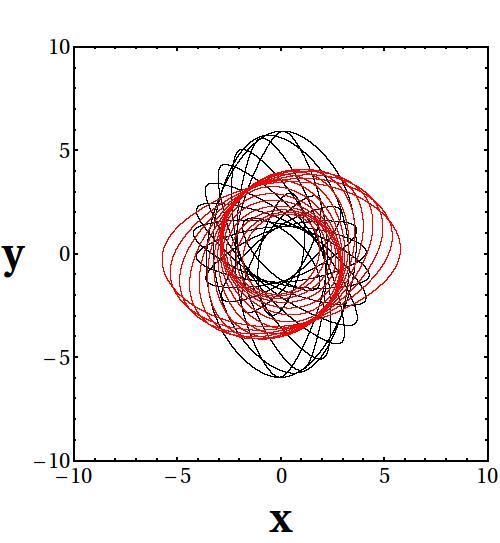}
\caption{Some precessing ellipses of the $x_1$ family (black orbits) together with the precessing ellipses of the $x_2$ family (red orbits) that correspond to the same radii $r_c$ for density of the spiral potential $\rho_0=30 \times 10^7 M_\odot~\rm{kpc}^{-3}$ and for a pattern speed  $\Omega_{\rm{sp}}$=15 $~\rm{km~s^{-1}~kpc^{-1}}$.  The main axes of the $x_2$ ellipses are perpendicular to the main axes of the $x_1$ family and do not support the spiral density wave. } \label{MW}
\end{figure}

In Fig. 15, some precessing ellipses of the $x_1$ family (black orbits) are plotted together with those of the $x_2$ family (red orbits) for $\rho_0=30 \times 10^7 M_\odot~\rm{kpc}^{-3}$ and $\Omega_{\rm{sp}}=15 ~\rm{km~ s^{-1}~ kpc^{-1}}$, in the range of radii from $r_c=2$ kpc to $r_c=5$ kpc.  The main axes of the ellipses of the $x_2$ family are perpendicular to the main axes of the $x_1$ family, and they exist for radii for which the amplitude of the spiral perturbation is close to zero (see Fig.1). Therefore they do not support the spiral density wave. It can easily be verified that the same is true for the unstable $x_3$ family  (not shown in the figures).

\subsubsection{Role of the pattern speed}

In order to examine the dependence of the response spiral density wave on the pattern speed $\Omega_{\rm{sp}}$, we fixed  the amplitude of the spiral perturbation to $\rho_0=5 \times 10^7 M_\odot~\rm{kpc}^{-3}$ and changed
the pattern speed of the spiral potential, comparing the cases $\Omega_{\rm{sp}}$=10, 15 and 20 $\rm{km~s^{-1}~kpc^{-1}}$.
Figure 16 shows the spiral density waves formed by the precessing ellipses of the $x_1$ family (black orbits)  for the pattern speed $\Omega_{\rm{sp}}=20~\rm{km~s^{-1}~kpc^{-1}}$ (in Fig. 16a) and $\Omega_{\rm{sp}}=10~ \rm{km~s^{-1}~kpc^{-1}}$ (in Fig. 16b). Superposed are the circles corresponding to the ILR radius (green circle), to the 4:1 resonance (blue circle), and to corotation (black circle).
By comparing Figs. 16a and b and Figs.7a and b, which all have the same amplitude $\rho_0=5 \times 10^7 M_\odot~\rm{kpc}^{-3}$, but different values of the pattern speed we derive the following conclusions. When the pattern speed decreases, (a) all the resonances are shifted outward (see Fig. 4), and therefore the spiral density waves reach larger radii. However, the ellipses become rounder when they approach the 4:1 resonance, and therefore the spiral density wave becomes less conspicuous at larger radii. (b) The region inside the first ILR becomes smaller, and the region between the first and second ILR increases. The elliptical orbits of the $x_1$ family in this latter region become much more elongated and intersect each other. (c) The width of the spiral arms grows with radius  (see Fig. 16b). Savchenko et al.  (2020) claimed that in the $86\%$ of a sample of 155 face-on grand-design spiral galaxies,  they observed that the arm width increases with radius (see Mosenkov et al. 2020, for an alternative interpretation of this phenomenon based on the mechanism of swing amplification).

\subsubsection{Role of the pitch angle}

\begin{figure}
\centering
\includegraphics[scale=0.35]{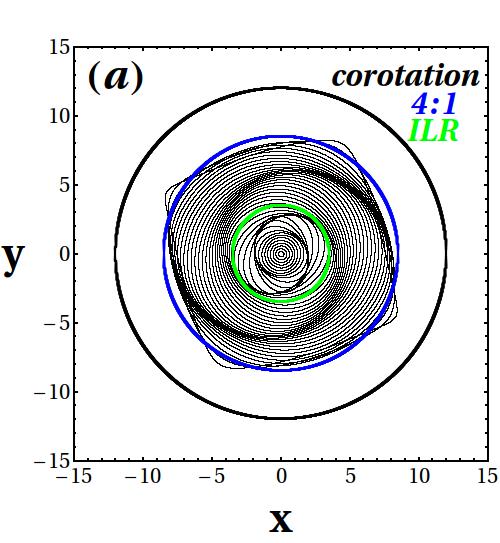}
\includegraphics[scale=0.35]{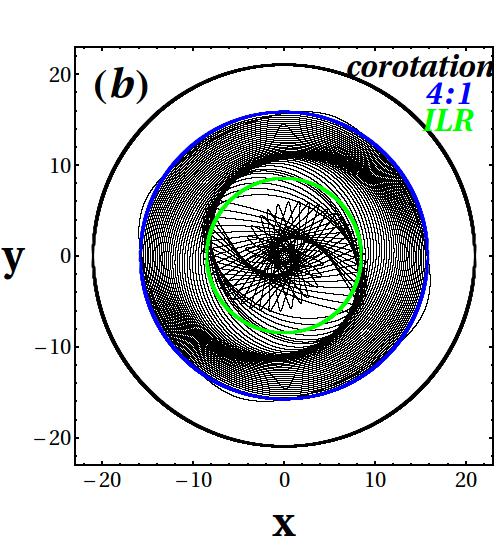}
\caption{Spiral density waves  formed by the elliptical orbits of the $x_1$ family, from the model of Eq. (8) for a pattern speed $\Omega_{\rm{sp}}=20~ \rm{km~s^{-1}~kpc^{-1}}$ in (a) and  $\Omega_{\rm{sp}}=10~ \rm{km~s^{-1}~kpc^{-1}}$ in (b) and a density of the spiral potential  $\rho_0=5 \times 10^7 M_\odot~\rm{kpc}^{-3}$. Superposed are circles corresponding to ILR (green circle), 4:1 resonance (blue circle), and  corotation (black circle).} 
\end{figure}
In this subsection, we chose the model $\rho_0=15 \times 10^7 M_\odot~ \rm{kpc}^{-3}$, $\Omega_{\rm{sp}}=15 ~\rm{km~s^{-1}~kpc^{-1}}$, but varied the pitch angle from the low value $a=-5^{\degree}$ to the high value $a=-25^{\degree}$ instead of the intermediate value 
$a=-13^{\degree}$ that was used in all previous numerical experiments.

\begin{figure*}
\centering
\includegraphics[scale=0.17]{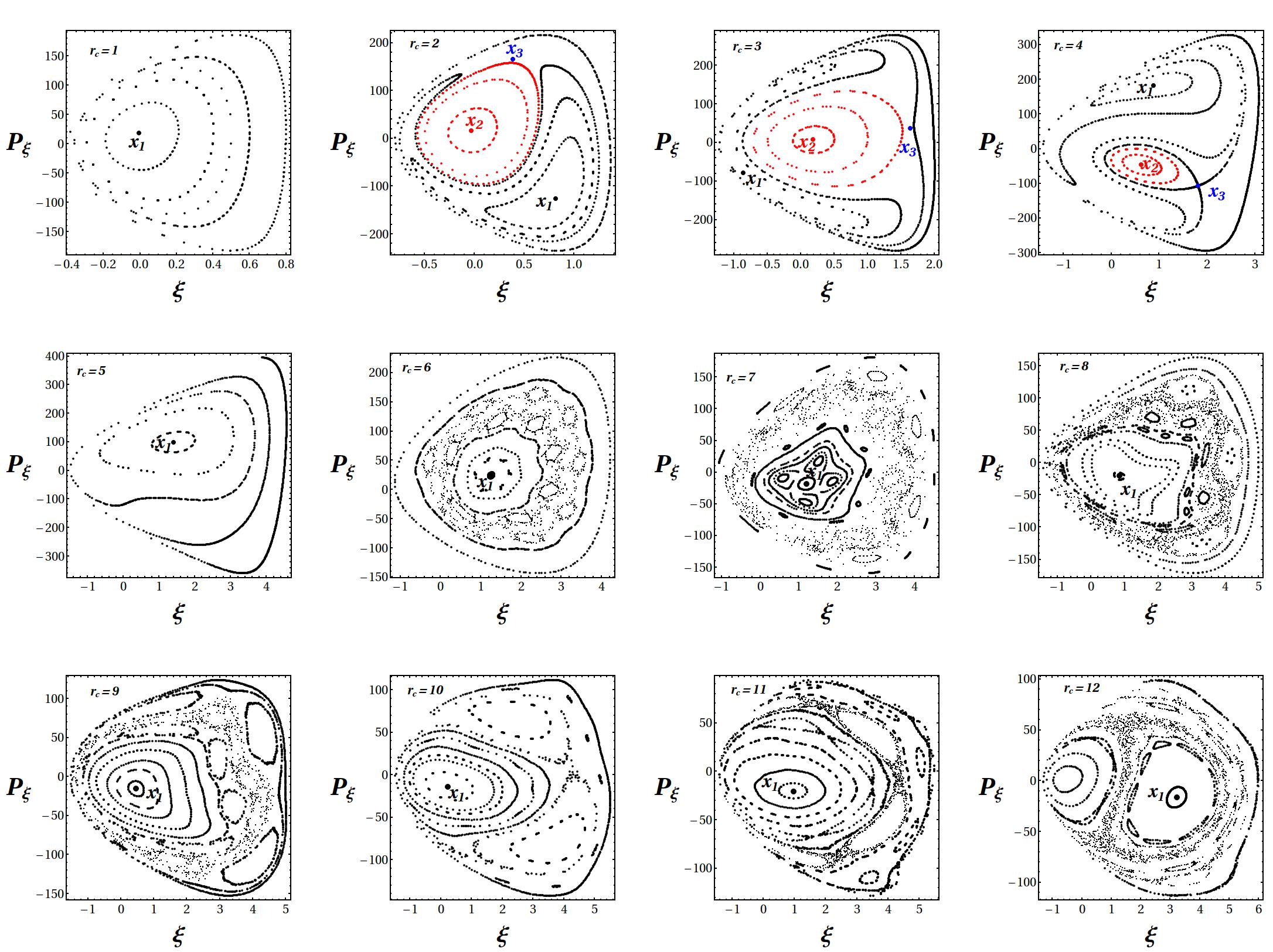}
\caption{Same as in Fig.8, but for  a pitch angle $a=-25^{\degree}$.} 
\end{figure*}
\begin{figure*}
\centering
\includegraphics[scale=0.17]{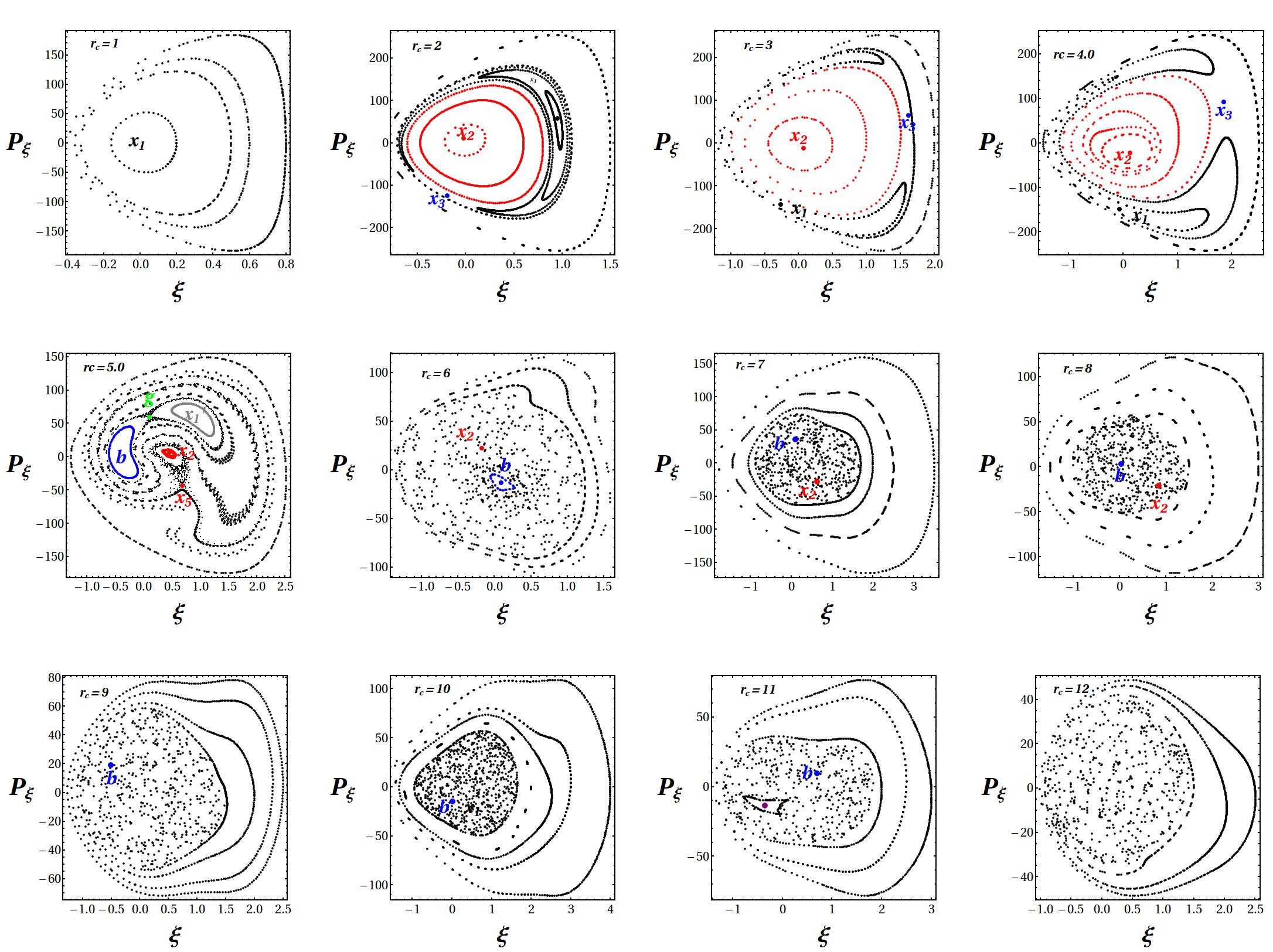}
\caption{Same as in Fig. 8, but for  a pitch angle $a=-5^{\degree}$.} 
\end{figure*}
\begin{figure}
\centering
\includegraphics[scale=0.38]{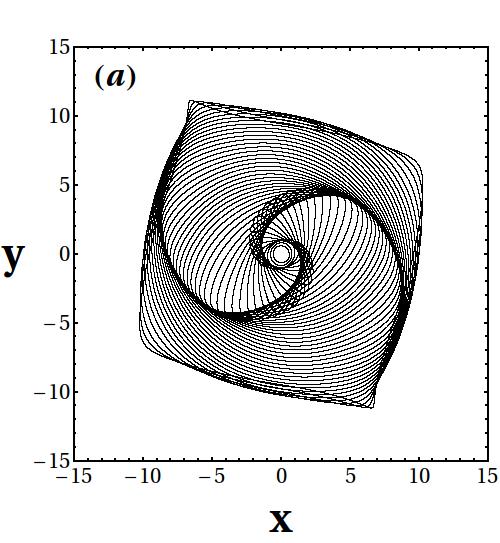}
\caption{ Spiral density wave formed by the precessing ellipses of the $x_1$ family for $\rho_0=15 \times 10^7 M_\odot~\rm{kpc}^{-3}$, a pattern speed $\Omega_{\rm{sp}}=15 ~\rm{km~s^{-1}~{kpc}^{-1}}$ , and a pitch angle $a=-25^{\degree}$. } 
\end{figure}
\begin{figure}
\centering
\includegraphics[scale=0.35]{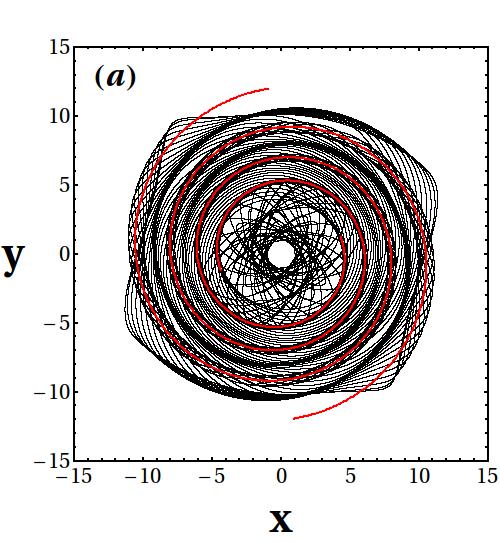}
\includegraphics[scale=0.35]{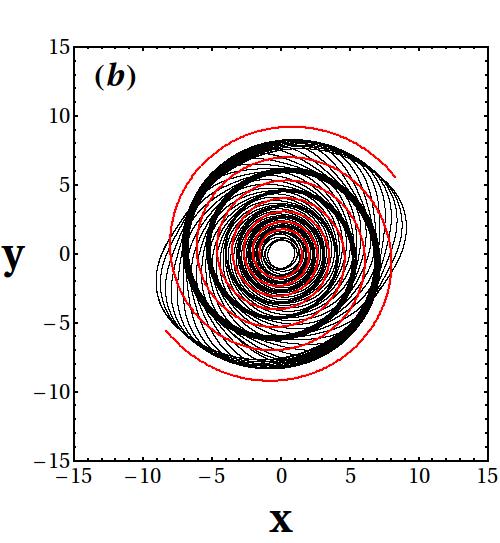}
\caption{ Spiral density waves formed by the precessing ellipses of the (a) $x_1$ family and (b) $x_2$ family for $\rho_0=15 \times 10^7 M_\odot~\rm{kpc}^{-3}$, a pattern speed $\Omega_{\rm{sp}}=15~ \rm{km~s^{-1}~kpc^{-1}}$ , and a pitch angle $a=-5^{\degree}$. Superposed (in red) is the theoretical spiral derived from the minima of the spiral potential of Eq. (8).} \label{MW}
\end{figure}
Figure 17 shows  the  phase portraits $(\xi, P_{\xi})$ for  $a=-25^{\degree}$. By comparing the two figures (8 and 17, which only differ in the value of the pitch angle), we conclude that for an increasing pitch angle (more open spiral arms), more order is introduced in the phase space and the chaotic areas shrink.
Therefore more matter exists in ordered motion around the $x_1$ stable periodic orbit that can support the spiral density wave better. The corresponding response spirals  (Fig. 19) are also more open and extend throughout the whole region between the second ILR and the 4:1 resonance.

Figure 18  instead shows the  phase portraits $(\xi, P_{\xi})$ when $a=-5^{\degree}$. The key observation is that the $x_1$ family now becomes unstable slightly outside the second ILR (at $r_c \approx 6.65 $kpc) and up to the 4:1 resonance, and the region around it is fully chaotic. This case presents an entirely different behavior than all previous cases. We studied the bifurcations of the $x_1$ and $x_2$ families in this case in the Appendix A in detail.  The corresponding spiral density wave formed by the $x_1$ family and its continuation, the $b$ family (of Appendix A), is plotted in Fig. 20a.  The ellipses of the $x_1$ family exist up to the second ILR (at $r_c  \approx 5$ kpc) and intersect, forming a rather fuzzy  density wave, while the ellipses from the $b$ family (which is the continuation of the $x_1$ family) create a more clearly defined density wave that reaches from the second ILR up to the 4:1 resonance (at $r_c \approx$ 11 kpc). Superposed, in red, are the theoretical minima of the imposed spirals. We observe that they coincide in general with the spiral density wave formed by the precessing ellipses of the $x_1$ family, although a secondary spiral wave appears from a certain radius and beyond. This is an  unrealistic spiral density wave as it is formed by unstable periodic orbits of the $x_1$ family that have no ordered matter around them, but only chaotic orbits all the way between the second ILR and the 4:1 resonance.  Figure 20b instead shows the spiral density wave formed by the $x_2$ family, which  in this model extends far beyond the second ILR (see the Appendix). As commented above, the main axes of the corresponding ellipses are always perpendicular to those of the $x_1$ family ellipses (see Fig. 15), and thus cannot support the imposed spiral arms.

By comparing Figs. 8, 17, and 18, we conclude that for an increasing pitch angle (more open spiral arms), more order is introduced in the phase space. For very small pitch angles, chaos dominates in the phase space, and we always obtain that the $x_1$ family of orbits becomes unstable almost immediately after the second ILR. The $x_2$ family is also unstable in the region outside
the second ILR.

In conclusion, there is a lower limit of the value of the pitch angle $a$ below which the central periodic orbit of the $x_1$ family becomes unstable in the region between the second ILR and the 4:1 resonance. This limit of the pitch angle is defined by the other two free parameters of the model, that is, by the amplitude of the spiral perturbation and the pattern speed of the spiral arm, as indicated in Table 1.
\begin{table}[h!]
  \begin{center}
    \caption{Pitch angle below which the $x_1$ family becomes unstable outside the second ILR for various values of the amplitude of the spiral perturbation and the pattern speed.}
    \label{tab:table1}
    \begin{tabular}{l|c|r} 
      \textbf{amplitude} & \textbf{pattern speed } & \textbf{pitch angle }\\
      $\rho_0$ & $\Omega_{\rm{sp}}$ & $a$ (in degrees) \\
      \hline
      5 & 10 & 1\\
      15 & 10 & 9\\
      30 & 10 & 16\\
       \hline
       5 & 15 & 1\\
      15 & 15 &5\\
      30 & 15 & 15\\
       \hline
       5 & 20 & 1\\
      15 & 20 & 4\\
      30 & 20 & 7\\
    \end{tabular}
  \end{center}
\end{table}

Table 1 can be used to estimate the permissible area of pitch angles that for $\rho_0$ and $\Omega_{\rm{sp}}$ as in the first two columns should be larger than the value reported in the third column as a function of the amplitude of the spirals and the pattern speed.  From these data, we conclude that using the stability of the $x_1$ family as a criterion,  a correlation between the pitch angle and the amplitude of the spiral perturbation suggests that spirals formed by precessing ellipses  should be stronger in amplitude when they are more open (larger $a$). Moreover, a correlation between the pitch angle and the pattern speed is indicated: the higher the value of the pattern speed, the more  tightly wound the spirals should be to maintain stability of the $x_1$ family.

\subsection{Comparison with observations}
In its classical version, the Hubble sequence for spiral galaxies implies that the bulge size and the  spiral arm winding should be highly correlated. According to this classification, the ``$\rm{Sa}$'' galaxies have tightly wound arms and fat nuclear bulges,
``$\rm{Sb}$'' galaxies have moderately wound arms and moderate nuclear bulges, and ``$\rm{Sc}$'' galaxies have loosely wound arms and small nuclear bulges. However, modern classification schemes for spiral galaxies imply a considerable departure from the classic Hubble sequence as regards the correlation between spiral arm winding type and bulge size.
Early studies suggested that spiral arms become tighter with increasing mass of the bulge (Morgan 1958, 1959, Kennicutt 1981, Bertin et al. 1989).
Furthermore, Seigar et al. (2005, 2006) reported a tight connection
between pitch angle and morphology of the galactic rotation
curve, quantified by the shear rate, with open arms associated
with rising rotation curves and tightly wound arms connected
to flat and falling rotation curves.
On the other hand, Hart et al. (2017)
 recently analyzed a large sample of galaxies selected from the Sloan
Digital Sky Survey (SDSS; York et al. 2000) and found very
weak correlations between pitch angle and galaxy mass, and a
surprising trend that the pitch angle increases with increasing
bulge-to-total mass ratio.
Yu and Ho (2019) found that the pitch angle decreases (arms are more tightly wound) in galaxies of earlier Hubble
type, more prominent bulges, higher concentration, and higher total galaxy stellar mass. However, there is a significant scatter in their measures.
Finally, Masters et al. (2019) and D\'{i}az-Garc\'{i}a et al. (2019) found little or no correlation between
spiral arm winding tightness and bulge size.  

There is no concluding evidence of a correlation between the size of the bulge and the pitch angle of the grand design galaxies. We therefore tested various pitch angles in our galactic model and kept the  mass of the bulge constant at a relatively high value. The results of  the previous subsection  indeed suggest a correlation between pitch angle and amplitude of the spiral perturbation, suggesting that galaxies with stronger perturbations should have more open spirals (greater value of the pitch angle).
In comparison with real observations, Grosb\o l et al. (2002) found that the distribution of mean amplitudes of two-armed spirals as a function of the pitch angle  shows a lack of strong in amplitude and at the same time tightly wound spirals. They have also found that most of the mean amplitudes of spiral arms are below $15\%$ in forces, and there is a correlation between the amplitude of spirals and the pitch angle: weak spirals often have tighter spiral arms. D\'{i}az-Garc\'{i}a et al. (2019) also found that the mean amplitude of the arms increases with increasing pitch angle.

\section{Conclusions}

We studied the precessing ellipse model of elliptical orbits that support the spiral density waves in grand-design spiral galaxies. We used a theoretical model consisting of a bulge, a disk, a halo, and a spiral potential. The free parameters of our model were the amplitude of the spiral density perturbation ($\rho_0$ in Eq. (8)), the pattern speed of the spiral potential ($\Omega_{\rm{sp}}$ in Eq. (12)), and the pitch angle of the spiral arms of the model ($a$ in Eq. (8)).
By testing the effect of the variation in free parameters of the model on the response spirals, formed by elliptical periodic orbits, we extracted the conclusions listed below.\\
1) In all models we studied, the $x_1$ family creates response spirals that are consistent with the imposed spirals. The response spirals extend  in the region from the center of the galaxy up to the radius of the 4:1 resonance. The $x_2$ and $x_3$ family of orbits exist between the first and second ILR. They are created simultaneously near the first ILR at a tangent bifurcation, and they join and disappear near the second ILR in all the models, except in the case of a very small pitch angle, where they still exist outside the second ILR, but do not contribute to the response spirals. The main axes of the $x_2$ family of orbits are perpendicular to the main axes of the $x_1$ family, and the $x_3$ family is always unstable in the whole range of radii.  \\
2) By increasing the amplitude of the spiral density perturbation in our model, chaos is introduced gradually and the $x_1$ family of orbits becomes unstable. An upper limit for the spiral density perturbation is $\rho_0=30 \times 10^7 M_\odot~\rm{kpc}^{-3}$ , where the $x_1$ family is unstable in the whole range between the second ILR and the 4:1 resonance.\\
3) When the value of the pattern speed $\Omega_{\rm{sp}}$ decreases, it causes an outward shift of the resonances and therefore the spiral density waves can reach larger radii. However, the ellipses become rounder when they approach the 4:1 resonance, and therefore the spiral density wave becomes less conspicuous at larger radii. Moreover, the elliptical orbits of the $x_1$ family become much more elongated and intersect, thus destroying a coherent spiral response.\\
4) We conclude for the value of the pitch angle that for an increasing pitch angle (more open spiral arms), more order is introduced in the phase space and the chaotic areas shrink, while for a decreasing pitch angle (more tight spiral arms), chaos dominates in the phase space and the $x_1$ family of orbits can no longer support the spiral density wave. In the special case of a very small pitch angle, new bifurcations of periodic orbits appear from the main families $x_1$ and $x_2$, but  the main bifurcations of the $x_1$ family that cause the spiral density wave become unstable soon after the second ILR, and therefore they cannot support real spirals.\\
5) To summarize, the main result of the paper is a quantitative estimation of the range of the free parameters of our galactic model for generating realistic spiral density waves through the precessing ellipses  model of elliptical closed orbits (and their surrounding quasi-periodic ones). In particular, a correlation between the pitch angle and the amplitude of the spiral perturbation was shown, where stronger spirals can be statistically less tight than weaker spirals. Moreover, a correlation between the pitch angle and the pattern speed was shown, where spirals that spin faster can be statistically tighter than slower spirals.\\

\begin{acknowledgements}
We acknowledge support by the research committee of the Academy of Athens through the project 200/895.
\end{acknowledgements}

\newpage

\textbf{References}\\
Athanassoula E.,  Phys. Rep., 114, 319, 1984\\
Berman R. H. and Mark J. W. K., ApJ, 216, 257,  1977\\
Bertin G., Lin, C. C., Lowe, S. A., and Thurstans R. P., ApJ, 338, 78, 1989\\
Berry C. L., and Smet, D. J.,  
Astron. J., 84, 964, 1979\\
Binney J. and Tremaine S., ``Galactic Dynamics'', Princeton University Press, 2008\\
Block D. L.,  Buta, R. , Knapen J. H., Elmegreen D. M., Elmegreen B.G., and Puerari I., Ap J, 128, 183, 2004 \\
Buta R. J., Knapen J. H., Elmegreen B. G., Salo H., Laurikainen, E., Elmegreen D. M., Puerari I., Block D. L.,  Astron. J., 137, 4487, 2009\\
Chakrabarti S., Laughlin G. and Shu F. H.,  ApJ, 596, 220, 2003\\
Chaves-Velasquez L,
Patsis P. A., Puerari I., Moreno E. and Pichardo B.,  ApJ, 871, 79, 2019\\
Contopoulos, G., in The Spiral Structure of Our Galaxy, ed. W. Becker and
G. Contopoulos, Proceedings of I.A U. Symposium No 38 (Dordrecht: D. Reidel
Publishing Co ), 1970\\
Contopoulos,  ApJ, 163, 181, 1971\\
Contopoulos G., ApJ, 201, 566, 1975 \\
Contopoulos G., J.,  A\&A, 1, 79, 1980\\
 Contopoulos G., Comments on Astroph., 11, 1, 1985\\
 Contopoulos G. and Grosb\o l P., A\&A, 155, 11, 1986\\
  Contopoulos G. and Grosb\o l P., A\&A, 197, 83, 1988\\
Contopoulos G., ``Order and chaos in dynamical astronomy'', Berlin, Springer, 2002\\
Cox, D. P., G\'omez G. C., ApJ Sup., 142, 261, 2002\\
 Dehnen, W., MNRAS, 265, 250, 1993\\
 D\'{i}az-Garc\'{i}a S., Salo H., Knapen J. H., Herrera-Endoqui M., A\&A, 631A, 94, 2019.\\
Dobbs C., Baba, J., Publ. Astron. Soc. Austr., 31, 35, 2014\\
Donner K. J. Thomasson M., A\&A, 290, 475, 1994\\
  Efthymiopoulos, Ch.,  Europ. Phys. J. Special Topics, 186, 91, 2010.\\
 Goldreich P. and Tremaine S., ApJ, 222, 850, 1978\\
 Grosb\o l, P. and Patsis P. A., A\&A 336, 840, 1998\\
 Grosb\o l P.,   Pompei E. and Patsis P.A., ASP conference series, 275, 305, 2002\\
 Grosb\o l P.,  Patsis P.A. and  Pompei E., A\&A 423, 849, 2004\\
 Hart R. E., Bamford S. P., Hayes W.B., MNRAS, 472, 2263, 2017\\
 Junqueira T. C., Lepine, J. R. D., Braga, C. A. S. and
Barros D. A., A\&A, 550, A91, 2013\\
 Kalnajs A.J.,  Proc. ASA, 2, 174, 1973\\
  Kennicutt R. C., Astron. J., 86, 1847, 1981\\
 Lepine J. R. D., Roman-Lopes A., Abraham Z., Junqueira T. C., Mishurov
Y. N., MNRAS, 414, 1607, 2011\\
Lin C. Shu F., ApJ, 140, 646, 1964\\
Lin C. Shu F., PNAS, 55, 229, 1966\\
Lindblad B., ApJ 92, 1, 1940\\
Lindblad B., Stockholm Obs. Ann. , 18, 6, 1955\\ 
Lindblad B.,  Stockholm Obs. Ann. , 19, 7, 1956\\
Lindblad B.,  Stockholm Obs. Ann. , 19, 9, 1957\\
Lindblad B., Stockholm Obs. Ann. ., 20, 4, 1958\\
Lindblad B., Stockholm Obs. Ann. , 21, 4, 1960\\
 Lindblad B., Stockholm Obs. Ann.  21, 8, 1961\\
  Lynden-Bell D. and Kalnajs A.J., MNRAS, 157, 1L, 1972\\
  Masters K.L., Lintott C.J., Hart R.E., Kruk S.J., Smethurst R.J., Casteels K.V.,   Keel W.C., Simmons B.D., Stanescu D.O.,  Tate J. and  Tomi S., MNRAS, 487, 1808, 2019\\
  Mertzanides, A\&A, 50, 395, 1976\\
  Miyamoto, M., Nagai, R., Publ. Astron. Soc. Japan, 27, 533, 1975\\
  Morgan, W. W., PASP, 70, 364, 1958\\
   Monet  D. G and Vandervoort,, ApJ, 221, 87, 1978\\
Morgan, W. W., PASP, 71, 394, 1959\\
Mosenkov A., Savchenko S. and  Marchuk A., Res. Astron. Astroph., 20, 120. 2020\\
Norman C. A., MNRAS, 182, 457, 1978\\
Patsis P. A., Contopoulos G., and Grosb\o l, P., A\&A, 243, 373, 1991\\
Patsis P. A., Hiotelis, N., Contopoulos G., and Grosb\o l P., A\&A, 243, 373, 1994.\\
Patsis P. A., Grosb\o l P., A\&A, 315, 371, 1996\\
Patsis P. A., Grosb\o l P., and Hiotelis N. , A\&A, 323, 762, 1997\\
Pérez-Villegas, A., Gómez, G. C., Pichardo B., MNRAS, 451, 2922, 2015 \\
Pettitt A. R., Dobbs C. L., Acreman D. M. and Price D. J., MNRAS, 444,
919, 2014\\
Pichardo B., Martos M., Moreno E., Espresate J., ApJ, 582, 230, 2003\\
Quillen A. C., Minchev I., ApJ, 130, 576, 2005\\
Savchenko S., Marchuk A., Mosenkov A., Grishunin K., MNRAS, 493, 390, 2020\\
Seigar M. S., Block D. L., Puerari I., Chorney N. E. and James, P. A.,  MNRAS, 359, 1065, 2005\\
Seigar M. S., Bullock J. S., Barth A. J. and Ho L. C., ApJ, 645, 1012, 2006\\
Sellwood, J. A. and Carlberg, R. G.,  ApJ, 282, 61, 1984\\
Sellwood J. A., in Gilmore G., ed., Planets Stars and Stellar Systems, Vol. 5. Springer, Heidelberg, preprint (arXiv:1006.4855), 2010\\
 Toomre A., ApJ, 139, 1217, 1964\\
 Toomre A., Ann. Rev.A\&A, 15, 437, 1977\\
 Tsigaridi L., Patsis P. A,  MNRAS, 434, 2922, 2013\\
 Vandervoort P. O., ApJ, 166, 37, 1971\\
 Vandervoort P. O.,  ApJ, 180, 739, 1973\\
 Vandervoort P. O. and Monet  D. G,  ApJ, 201, 311, 1975\\ 
 York, D. G., Adelman, J., Anderson, J. E., A\&A, 120, 1579, 2000\\
  Yu S. and Ho L., ApJ, 871, 194, 2019\\
Zhang X., ``Dynamical evolution of galaxies'', De Gruyter, GmbH, Berlin/Boston, 2018\\

\begin{appendix}

\section{Bifurcations of the orbits in the case of a small pitch angle}

Section 4.2.3 describes the case of a very small pitch angle, $a=-5^{\degree}$, for the amplitude of the spiral perturbation $\rho_0=15 \times 10^7 M_\odot~\rm{kpc}^{-3}$  and  a pattern speed $\Omega_{\rm{sp}}$=15$~ \rm{km~s^{-1}~kpc^{-1}}$ in our galactic model. In this case, the families of periodic orbits emerge from an intricate sequence of bifurcations. In this appendix, we present some further details of these families.

Figure 18 shows the phase-space portraits of these families for integer values of the radius of the circular orbit $r_c$, namely $r_c=1,2,3,..12$ kpc. The various bifurcations appear at values of $r_c$ between those shown in the figure.

Figure A.1 shows the characteristic curves $S(r_c)=\sqrt{\xi^2+p_{\xi}^2/\kappa_c ^2}$ of  the various families for this case. Fig. A.1.a shows the  characteristic curve of  the $x_1$ family and its bifurcations, while Fig. A.1.b shows the  characteristic curve of  the $x_2$ family and its bifurcations, as well as the $x_3$ family.
Figure A.2 shows the phase-space portraits for radii $r_c=4.4, 4.5, 4.6, 4.7, 4.78, 4.8, 5.0,$ and $5.5$ kpc ,which give  a more detailed description of Fig. 18 in the range of radii between $r_c=4$ and $r_c=5.5$, where the bifurcations of the $x_1$ and $x_2$ families take place. 
\begin{figure}
\centering
\includegraphics[scale=0.23]{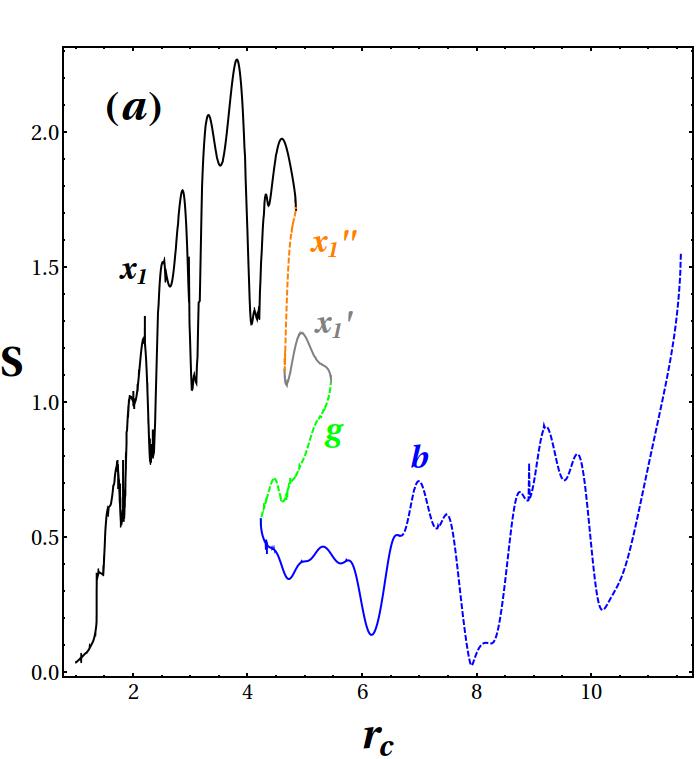}
\includegraphics[scale=0.23]{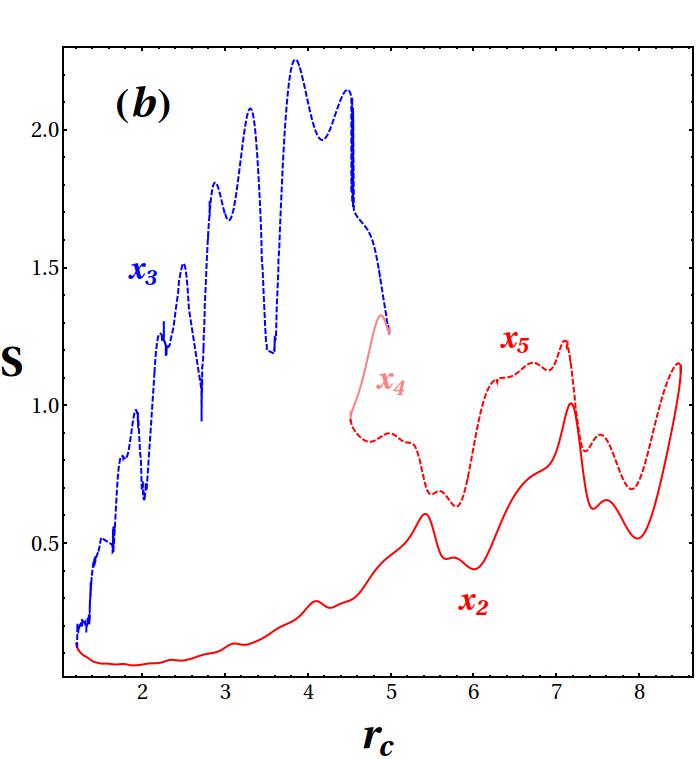}
\caption{(a) Characteristic curves $S(r_c)=\sqrt{\xi^2+p_{\xi}^2/\kappa_c ^2}$ of the periodic families $x_1$ (black) and their bifurcations $x_1'$ (gray),  $x_1''$ (orange), $g$ (green), and $b$ (blue). The dashed part of the blue curve denotes that the $b$ family is unstable. (b) The characteristic curves $S(r_c)$ of the periodic families $x_3$ (dashed blue) and $x_2$ (red), and their bifurcations $x_4$ (pink) and  $x_5$ (dashed red). The parameters of the model of Eq. (8) are $\Omega_{\rm{sp}}=15 ~\rm{km~s^{-1}~kpc^{-1}}$, $\rho_0=15 \times 10^7 M_\odot~\rm{kpc}^{-3}$ , and $a=-5^{\degree}$.}
\end{figure}
\begin{figure*}
\centering
\includegraphics[scale=0.17]{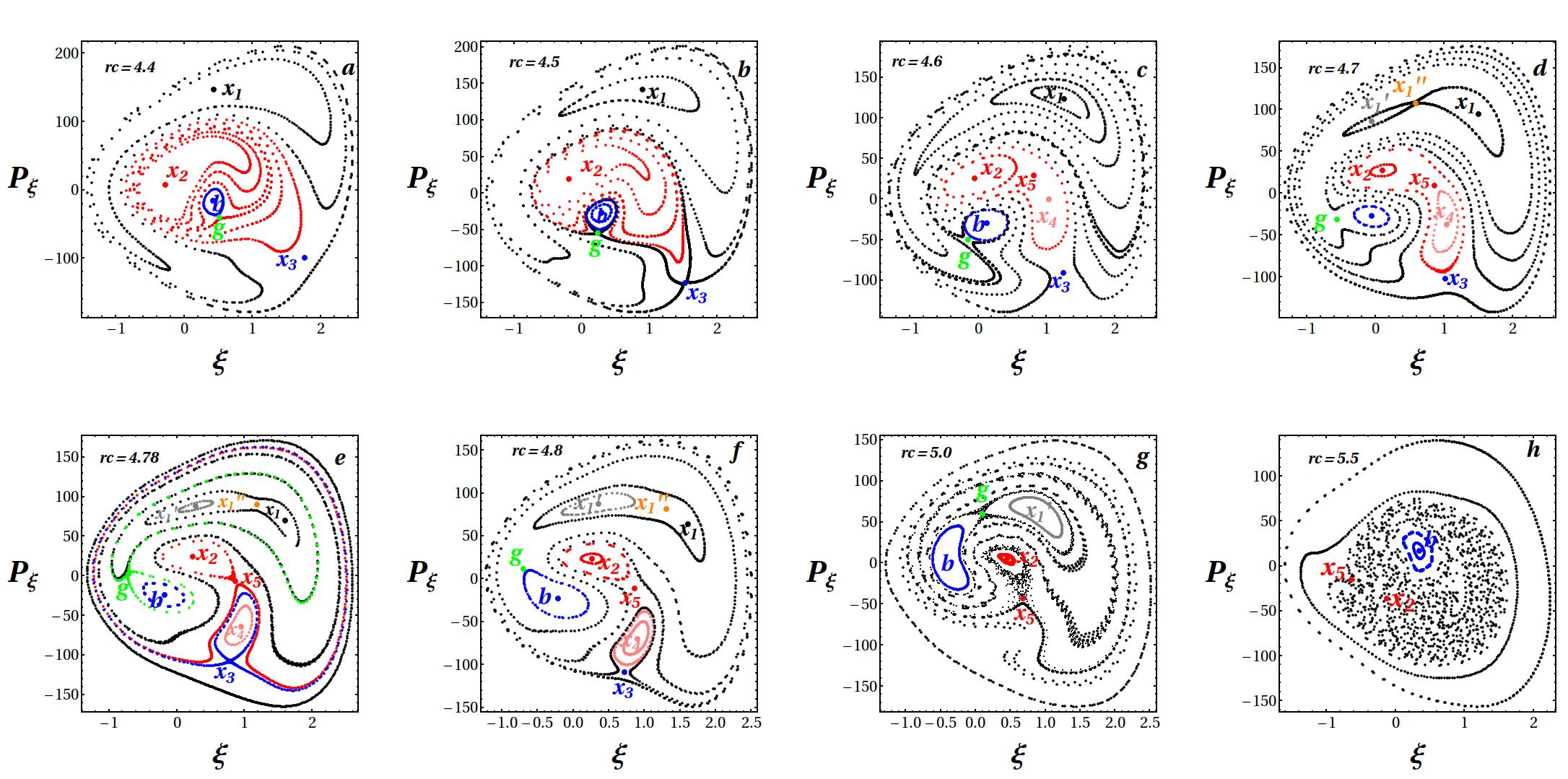}
\caption{Phase-space portraits $(\xi, P_{\xi})$ for the model of Eq. (8) with pattern speed $\Omega_{\rm{sp}}=15 ~\rm{km~s^{-1}~kpc^{-1}}$,   $\rho_0=15 \times 10^7 M_\odot~\rm{kpc}^{-3}$ , and $a=-5^{\degree}$, for values of the radius $r_c$ =4.4, 4.5, 4.6, 4.7, 4.78, 4,8, 5.0, and 5.5 kpc.   Precessing ellipses responsible for the spiral density waves are periodic orbits of the $b$ family (bifurcation of the $x_1$ family).}
\end{figure*}
These figures show that
a) for values of $r_c$ close to zero, only the stable $x_1$ family exists. This family exists all the way up to $r_c \approx 4.84$, where it joins the unstable family $x_1''$ and disappears.
b) The families $x_2$ (stable) and $x_3$ (unstable) are generated at a tangent bifurcation at $r_c \approx 1.2$. The family $x_2$ exists all the way up to $r_c \approx 8.5$, where it joins the unstable family $x_5$ and disappears. The family $x_3$ exists up to $r_c \approx 4.98,$ where it joins the unstable family $x_4$ and disappears.
c) The phase-space portrait $(\xi,P_{\xi})$ for $r_c=4.4$ (Fig. A.2.a) shows $x_1$, $x_2$, $x_3$ , and some new bifurcated orbits  inside the domain covered by red invariant curves, surrounding the $x_2$ periodic orbit. These orbits are called $b$ (blue stable periodic orbit) and $g$ (green unstable periodic orbit) and have been generated at $r_c \approx 4.23$ at a tangent bifurcation (Fig. A.1.a).  Fig. A.2.a also clearly shows that the separatrix from $x_3$ surrounds $x_2$ (inner curve of separatrix) and  both $x_1$ and $x_2$ (outer curve of separatrix).
d) From Fig. A.2.b ($r_c=4.5$), we conclude that for a slightly smaller $r_c$ , the unstable periodic orbit $g$ has moved downward and has reached the separatrix emanating from $x_3$. e) Fig. A.2.c ($r_c=4.6$) shows that the  separatrix of $g$ surrounds both the $x_1$ and the $b$ periodic orbits. The separatrix of $x_3$ no longer surrounds the orbit $b$ .
f) For $r_c=4.6$ (Fig. A.2.c), two new orbits appear in the domain around the $x_2$ periodic orbit, namely orbits $x_4$ (stable) and $x_5$ (unstable) created at a tangent bifurcation at $r_c \approx 4.51$ (Fig. A.1.b). The separatrix emanating from $x_5$ surrounds both $x_2$ and $x_4$. Moreover, the separatrix of $x_3$ surrounds all the three orbits, namely $x_2$, $x_4$ , and $x_5$.
g) Fig. A.2.d ($r_c=4.7$) shows that two more families of orbits are created inside the region surrounding the $x_1$ periodic orbit, namely $x_1'$ (stable) and $x_1''$ (unstable). These families are generated  at a tangent bifurcation  at $r_c \approx 4.64$ (Fig. A.1.a).
h) Fig. A.2.e ($r_c=4.78$) shows that the separatrix emanating from $x_3$ has almost reached the unstable orbit $x_5$ and the separatrix emanating from $x_5$ (that surrounds $x_4$) has almost reached $x_3$.   For a slightly smaller value of $r_c$ , these parts of the two separatrices should coincide.
i) At $r_c = 4.8$  (Fig. A.2.f), the separatrix emanating from $x_3$ surrounds only $x_4$. On the other hand, the separatrix emanating from $x_5$ surrounds on the one side the periodic orbit $x_2$ and the other side all the periodic orbits $x_1, x_1', x_1'', b,$ and $g$.
j) At $r_c \approx 4.98,$  the orbits $x_3$ and $x_4$ join and disappear, and the same happens for orbits $x_1$ and $x_1''$ at $r_c \approx 4.84$ (Figs. A.1.a,b). Thus, for $r_c=5$ (Fig. A.2.g), only the periodic orbits $x_1'$ (stable), $b$ (stable), $g$ (stable), $x_2$ (stable), and $x_5$ (unstable) exist.
k) At $r_c\approx 5.45,$ the orbits $x_1'$ and $g$ join and vanish (Fig. A.1.a). Thus, for $r_c=5.5$ (Fig. A.2.h), only orbits $b$ (stable), $x_2$ (stable), and $x_5$ (unstable) remain.
l) At $r_c \approx 8.5$, families $x_2$ and $x_5$ join and vanish (Fig. A.1.b), and finally, beyond $r_c \approx 8.5$ only the $b$ periodic orbit remains up to the 4:1 resonance. This is the main family that could support the spiral density wave, but it is unstable in most of the domain between
the second ILR and the 4:1 resonance. Although we have constructed a density wave out of the precessing ellipses of the unstable periodic orbit $b$ (see Fig. 20a), the fact that there are no ordered orbits around this periodic orbit, has as a result that such a spiral density wave cannot be observed in real galaxies.

The sequence of bifurcations that we observe in Figs. A.1.a and b is much more complicated than in the corresponding Figs. 6, 9, and 12, where only families $x_1$, $x_2$ and $x_3$ exist. In these cases, families $x_2$ and $x_3$ start at a tangent bifurcation and disappear by joining again. Family $x_1$ is stable  all the way to the 4:1 resonance and beyond it for a small amplitude of the spiral perturbation ($\rho_0=5 \times 10^7 M_\odot~\rm{kpc}^{-3}$, Fig. 4), but it becomes unstable at a certain radius between the second ILR and the 4:1 resonance by a sequence of period-doubling bifurcations for higher values of the amplitudes, that is, for  $\rho_0 \geqslant 15 \times 10^7 M_\odot~\rm{kpc}^{-3}$ (see Figs. 8 and 11).

In all these cases, the pitch angle is $a=-13^{\degree}$. For larger pitch angles (see Fig. 16 for $a=-25^{\degree}$), the $x_1$ family is stable at least up to the 4:1 resonance.

In the present case, where the pitch angle is
significantly smaller, that is, $a=-5^{\degree}$, family $x_1$ has a very different evolution.
The characteristic curve of $x_1$ (stable) is continued by the characteristic curve of $x_1''$ (unstable and backward in $r_c$), then $x_1'$ (stable and forward in $r_c$), $g$ (unstable and backward in $r_c$), and $b$ (stable and forward in $r_c$) (Fig. A.1.a), which finally becomes unstable by a period-doubling bifurcation slightly after  $r_c \approx 6$ and produces great chaos. On the other hand, the characteristic curve of $x_3$  (unstable) starts at tangent  bifurcation with $x_2$ (stable)  and continues with the characteristic curve of $x_4$ (stable, backward in $r_c$), the $x_5$ (unstable), and finally, it joins $x_2$ (stable) at $r_c \approx 8.5$ and disappears.  Thus the extended characteristic curves of $x_3$ and $x_2$ have the same starting and ending points as in all the previous models.

Another strange new feature in this model is the evolution of the separatrices of the new unstable families  $g$ and $x_5$. In both cases, the unstable points of $g$ and $x_5$ reach the separatrix emanating from the unstable point $x_3$ at critical values of $r_c=r_{c_{crit}}$. After this value of $r_c$ , the separatrices emanating from $g$ or $x_5$ change drastically. 

\end{appendix}

\end{document}